# Electron Photodetachment from Aqueous Anions.

# I. Quantum Yields for Generation of Hydrated Electron by 193 and 248 nm Laser Photoexcitation of Miscellaneous Inorganic Anions. [1]

Myran C. Sauer, Jr., R. A. Crowell, and Ilya A. Shkrob *

*Chemistry Division, Argonne National Laboratory, Argonne, IL 60439*

**Abstract**

Time resolved transient absorption spectroscopy has been used to determine quantum yields for electron photodetachment in 193 nm and (where possible) 248 nm laser excitation of miscellaneous aqueous anions, including hexacyanoferrate(II), sulfate, halide anions ($Cl^-$, $Br^-$, and $I^-$), pseudohalide anions ($OH^-$, $HS^-$, $CNS^-$), and several common inorganic anions for which no quantum yields have been reported heretofore: $SO_3^{2-}$, $NO_2^-$, $NO_3^-$, $ClO_3^-$ and $ClO_4^-$. Molar extinction coefficients for these anions and photoproducts of electron detachment from these anions at the excitation wavelengths were also determined. These results are discussed in the context of recent ultrafast kinetic studies and compared with the previous data obtained by product analyses. We suggest using electron photodetachment from the aqueous halide and pseudohalide anions as actinometric standard for time-resolved studies of aqueous photosystems in the UV.

---

## 1. Introduction.

Photoinduced electron detachment from aqueous anions, such as halides and pseudohalides, is a rapid, efficient charge transfer reaction in which the solvent itself serves as an acceptor of the ejected electron (charge transfer to the solvent, CTTS). [1,2] For any photoreaction, the quantum yield (QY) is one of the most important photophysical parameter, and electron detachment from the anions is no exception to this rule. While many estimates for the quantum yields of hydrated electron generated in the course of CTTS photoreactions have been given in the literature (section 1.1), these estimates agree poorly with each other and strongly depend on the actinometric standards used. Furthermore, recent kinetic data (discussed below) [3-14] suggest that the very approach used to obtain these quantum yields 20-40 years ago, by product analysis, was inherently inconsistent (as was realized by several authors in the 1960s [15,16]). On the other hand, accurate QY measurements are necessary for understanding the photophysics of charge separation on short time scales [4]. In this work, we report quantum yields and cross sections for fifteen aqueous anions photoexcited at 193 and 248 nm. Due to unexpectedly complex photophysics and chemistry, these measurements turned out to be rather involved. This article deals mainly with these QY measurements *per se* and various complications encountered along the way. A meaningful interpretation of these QY data is possible only when these QY data on the free electron yield are combined with ultrafast kinetic measurements (as done in ref. [4] for iodide); this task is deferred to another publication (Part III of this series [17]).

### *1.1. Background and Motivation.*

In polar liquids, including aqueous solutions, some anions ($X^-$) exhibit absorption bands (CTTS bands) that are lacking in their gas phase spectra. Photoexcitation of these anions in their CTTS bands causes rapid (< 150 fs [5]), efficient (with the prompt QY approaching unity [4]) detachment of the electron from the photoexcited anion to the solvent. Solvated electrons generated in this reaction are observed on the time scale of $10^{-11}$ to $10^{-5}$ s [3-7,12-14]. Because of the fundamental importance and perceived simplicity of this CTTS photoreaction, much effort has been devoted to studying the photophysics of electron detachment from halide anions. In particular, there have been several ultrafast studies in which the electron dynamics were studied on pico- [3-7] and femto- [5] second time scales. These studies were complemented by state-of-the-art molecular dynamics [18,19,20] and *ab initio* calculations [2] that shed light on the nature of the short-lived CTTS state and mechanistic aspects of the charge separation. Several other aqueous anions have been studied in addition to iodide, most significantly, ferrocyanide (hexacyanoferrate(II)) [13,14], which is a common source of hydrated electrons in photochemical studies. Interestingly, this polyvalent anion exhibited electron dynamics that were entirely different from those observed from the halide and pseudohalide anions.

For the majority of aqueous anions, fast kinetic studies of one-photon CTTS reactions were slow to come, because no convenient source of pulsed excitation in the UV (where these anions have their CTTS bands) was available. Recently, we used short 200 nm pulses (300 fs fwhm) to study the electron dynamics for several aqueous anions, [12,17,21] including halides (I$^-$ and Br$^-$), pseudohalides (HO$^-$, HS$^-$, and CNS$^-$), and divalent anions (SO$_3^{2-}$, CO$_3^{2-}$). These 200 nm studies, and complementary 250-220 nm studies carried out by Bradforth and coworkers [3-7] indicate the complexity of photoexcitation process in the CTTS systems. From these studies it appears that two different mechanisms operating in tandem are responsible for the electron detachment from the aqueous anions:

The first mechanism involves the dissociation of a short-lived CTTS state and results in a narrow distribution of the electrons around the parent species (X• radical or atom). Both the kinetic studies [4,12,17] and molecular dynamics simulations [20] suggest the existence of a short-scale, attractive mean force potential between these electrons and the residual species (at least, for halides and pseudohalides). The resulting kinetics exhibit a fast (10-50 ps) exponential component due to the recombination and escape of the electrons residing near the bottom of the mean force potential well. These fast kinetics are succeeded by a long-term $t^{-1/2}$ decay on subnanosecond (and longer) time scales; [4,12,22] this decay is due to diffusional migration and recombination of the electrons that are thermally emitted from the potential well to the solvent bulk.

The second mechanism by which the hydrated electrons are generated is direct ionization of the anion by the UV photons. This competing process results in the formation of a mobile conduction band electron. This energetic electron localizes and thermalizes far from the parent species (> 1 nm), and the resulting geminate decay kinetics are purely diffusional. According to our studies [17] and work by Bradforth and coworkers [13,14], polyvalent anions yield this type of kinetics exclusively. Some polyatomic monovalent anions (e.g., thiocyanate) also seem to yield these kinetics only, even at low photoexcitation energy [17]. For other monovalent anions, the direct ionization may or may not occur, depending on the excitation energy [10,11,17]. E.g., for iodide, photoexcitation using light with wavelength > 220 nm yields decay kinetics that are independent of the excitation wavelength; however, 200 nm kinetics are very different from these 220-250 nm kinetics. [11,17] In particular, the 200 nm kinetics exhibit higher escape yield of the electron and demonstrate other features that are consistent with broadening of the electron distribution around the iodine atom [17]. Using 2-photon excitation, Bradforth and coworkers [11] irradiated aqueous iodide at progressively higher energy and found that the escape yield systematically increased with the total excitation energy. These observations suggest a competition between the direct ionization and the CTTS process. Such a competition has been invoked previously to account for the observed increase in the quantum yield for electron detachment with increase in photoexcitation energy (e.g. refs. [23] and [24]).

Most of the photophysical studies on anion CTTS were carried out in the 1960s, before the complexity of the electron dynamics was recognized. With few exceptions, the approach was to convert electrons to $N_2$ using 10 mM $N_2O$ as a scavenger

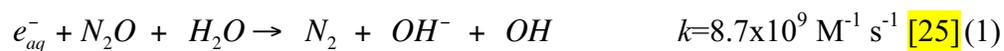

$$e_{aq}^- + N_2O + H_2O \rightarrow N_2 + OH^- + OH \qquad k=8.7\times10^9 \text{ M}^{-1}\text{ s}^{-1} \text{ [25]} \quad (1)$$

and measure the evolution of gaseous $N_2$. High concentration of the scavenger was needed to prevent the loss of photogenerated electrons to recombination [15,16]. As a result, some of the electrons were scavenged geminately, and this was recognized by Airey and Dainton (in their studies of ferrocyanide) [15] and Jortner et al. (in their studies of iodide) [23]. Picosecond kinetic studies justify the concerns of these authors since retrospective analysis of their data reveals many inconsistencies originating in the complexity of the electron dynamics [4]. While short-term exponential kinetics are too rapid to compete with the relatively slow reaction (1) [4,12], long-term kinetics unfolding after this fast exponential decay are sufficiently slow, and some geminate electrons are scavenged by $N_2O$. Thus, a QY for $N_2$ evolution via reaction (1) provides an *upper limit* estimate for the QY of free electron formation. Since the relative weight of the fast (exponential) and slow (power law) kinetics varies for different photosystems (e.g., refs. [4], [13], and [17]) and depends on the relative efficiency of direct ionization and CTTS excitation for a given photon energy, [10,11,17] even qualitative comparison between different photosystems is questionable.

Airey and Dainton [15] pointed out to another consequence of the geminate scavenging in $N_2O$-saturated aqueous solutions: In reaction (1), (geminate) electron is converted to a (geminate) hydroxyl radical. This radical rapidly reacts with the residual species X•, changing the product distribution. E.g., geminate OH radical can react with the cyanide ligand of ferricyanide that is formed in electron photodetachment from aqueous ferrocyanide; the resulting CN radical converts ferricyanide (hexacyanoferrate(III)) back to ferrocyanide [15]. Consequently, the QY of the electron estimated from the ferricyanide yield is always *underestimated*. This conclusion pertains to other measurements in which the QY for the electron was estimated from the yield of a stable photoproduct derived from X• (e.g., $I_3^-$ in the photoexcitation of iodide [26] and ferricyanide in the photoexcitation of ferrocyanide [27]).

For higher excitation energies (180-200 nm), an additional source of uncertainty is actinometry. Dainton and Fowles [28,29] based their measurements for $OH^-$, $SO_4^{2-}$, $Cl^-$, and $Br^-$ [29] on the quantum yield of $H_2$ (0.3) in 185 nm photolysis of 0.01 M aqueous methanol (which, in turn, was estimated from the QYs of 0.3 and 0.46 for neat water and methanol, respectively, and the assumed QY of unity for photodissociation of aqueous $N_2O$), [28] whereas Jortner et al. assumed a QY of 0.65 for the same photosystem [30]. Such a discrepancy exists for other "standard" actinometers for the UV. E.g., the estimates for the QY of $H_2$ generation in the ethanol-water actinometer at 185 nm range from 0.4 to 0.8 [31].

It is apparent from the above that for the QY measurement to be reliable, (i) the electron yield should be determined directly, on a fast time scale, and (ii) precision photometry rather than actinometry should be used to determine the fluence of the UV photons. To our knowledge, only one previous study met these requirements. Iwata et al. [24] used pulsed UV light from an excimer laser and measured the absorbance of hydrated electron at 720 nm, ca. 50 ns after the end of the laser pulse. Photooxidation of water by $Eu^{3+}$ was used for actinometry at 193 nm [32]. For $OH^-$, $SO_4^{2-}$, $Cl^-$, and $Br^-$, the 193 nm measurement gave QYs that were not too far from the 185 nm yields obtained by Dainton and Fowles; [29] however, for $I^-$, it gave a QY that was quite different from that obtained by Jortner et al. [30]

In this work, we improve on these time-resolved measurements and extend the method to less common CTTS (and suspected CTTS) anions (such as carbonate, bicarbonate, nitrate, nitrite, chlorate, and perchlorate). We also give one-photon quantum yields for ionization of neat water by 193 nm light and dissociation of aqueous hydrogen peroxide by 248 nm light. In the course of this study, we realized that the QY data obtained by Iwata et al. [24] were compromised due to the absorption of the laser light by photoproducts generated within the duration of the laser pulse; our analysis includes corrections for this complicating effect. The correction is largest for anions that are efficient electron donors but absorb the excitation light poorly (e.g., $SO_3^{2-}$ at 248 nm; Table 2) and anions that yield products that strongly absorb at the excitation wavelength (e.g., ferricyanide at 193 nm; Table 2). The absorptivity of these photoproducts can be estimated from the power dependence of the laser light transmittance, and we give such estimates for several species, including hydrated iodine atoms. Other complications included secondary chemistry, reactions of electrons with impurity (e.g., for hydrosulfide), competition between mono- and bi- photonic excitation (for $ClO_4^-$), protic equilibria (for $HS^-$ and $CO_3^{2-}$), ion pairing (for ferrocyanide), and the dependence of the QYs on the ionic strength of (concentrated) solutions. Since many of these anions exhibit one peculiarity or another, they are dealt with on a case to case basis (section 4). This survey is preceded by the analysis of pertinent photophysics given in section 3. To save space, some figures (Figs. 1S to 12S) are given in the Supporting Information.

## 2. Experimental.

*The setup.* Unlike Iwata et al., [24,32] who used 3 cm optical path cell and 90° detection at the cell midsection (where laser light was attenuated by the sample), a short optical path ($L$=1.36 mm) cell and 30° detection were used in this work, i.e., all the electrons generated along the path of the excitation light were probed with the analyzing light. Another difference is our use of photometry instead of actinometry and direct measurement of the absorbed and transmitted laser power. These modifications allowed us to deal with the complications introduced by the absorption of the laser light by photoreaction products.

Fifteen nanosecond fwhm, 1-20 mJ pulses from an ArF (193 nm) or KrF (248 nm) excimer laser (Lamda Physik LPX 120i) were used to photolyze $N_2$-saturated aqueous solutions of the anions. The optical path cell had detachable 1 mm thick suprasil windows sealed to the stainless steel body of the cell with parafilm gaskets. Following the suggestion of Iwata et al., [24] we used $CaF_2$ windows in some of the initial experiments; however, this material was found to be unsuitable due to rapid generation of color centers by 193 nm laser light and corrosion of these windows by photoexcited ions (such as sodium thiocyanate). The refractive index for Suprasil is 1.508 at 248 nm and 1.56 at 193 nm, [33] and the calculated reflectivity of the window is 0.048 and 0.04, respectively. Typical transmittance of these windows at 193 nm was 90.6 to 91.5%. The reflection losses at the glass-window boundary were factored into the QY calculation; the reflection losses at the glass-water boundary were ignored due to the similarity of the refraction indexes for the two media (for water, the refraction index is 1.4 at 193 nm and 1.35 at 248 nm, [34] so the reflection losses are < 0.1%).

The laser beam was focused by a 50 cm focal length, 3 mm thick $CaF_2$ lens and uniformly illuminated a rectangular 3 mm x 6 mm, 130 μm thick brass aperture placed onto the front window. This laser beam was normal to the window. The beam energy was attenuated using a set of fine wire mesh filters placed 1.5 m away from the focussing lens (the pattern cast by these filters was destroyed by diffraction). The analyzing light from a superpulsed 75 W Xe lamp was passed through 4 cm of water (to reduce heat transfer to the sample), a color glass filter (> 500 nm), and crossed at 30º with the excitation beam. After traversing the sample, the analyzing light passed through another color glass filter and was focussed on the detector using a 7 cm focal length achromat. The wavelength of the analyzing light (typically, 700 nm) was selected using a 10 nm fwhm band interference filter. With our beam and cell geometry, ca. 18% of the cell volume illuminated by the laser light was not probed by the analyzing light, and the effective mean path for the analyzing light was 1.29 mm. A correction was made to take this shadowing into the account for the QY calculations.

A fast silicon photodiode (EG&G model FND100Q, biased at -90 V) with a 1.2 GHz video amplifier (Comlinear model CLC449) terminated into a digital signal analyzer (Tektronix model DSA601) were used to sample the transient absorbance kinetics (3 ns response time). Two calibrated, NIST traceable, lithium tantalite pyroelectric energy meters (Molectron model J25-080 with black oxide coating that provides flat spectral response in the UV) were used to measure the power of the incident and transmitted UV light. To this end, a small fraction of the excitation light was diverted to one of the meters using a thin suprasil beam splitter before the sample cell; the second meter was placed ca. 50 cm behind the cell (where the beam is sufficiently expanded). The output signals from these energy meters were amplified using Gentec PRJ-D meters, terminated into 1 MΩ and sampled at 200 kHz using a 12-bit ADC board (National Instruments PCI-6064E). 8-to-10 of these digitized waveforms were averaged and the peak signals were converted to the laser energy using calibration values provided by the manufacturer. A series of 6-10 such measurements were collected and the energies given by the two detectors linearly correlated, in order to determine relative transmission of the laser light through the sample (this averaging and analysis was performed during kinetic sampling). Typical standard errors of the laser power and transmission measurements were 1-3% and 0.1-0.7%, respectively. Two power measurements, one for the cell filled with pure water and one for the aqueous solution being studied, were taken for each laser power. The difference between the measurements of transmitted energies (corrected by the window reflectivity) gave the number of absorbed photons, whereas their ratio (corrected by the ratio of incident laser energies) gave the transmission coefficient. To assess the reliability of the absolute power measurements, the readings from J25-080 pyroelectric detector were compared with the readings from a thermopile detector (Gentec ED500 with EM-1 meter). These two detectors were calibrated at 248 and 1064 nm, respectively. When the latter is corrected by 6% to take into account the reflectivity of the coating at 248 nm, the two readings were within 2% of each other.

*Materials and the flow system.* 0.2-0.5 L of aqueous solution of the salts was circulated through the cell using a peristaltic pump. The typical flow rate was 2-3 ml/min; the repetition rate of the laser was 1.7 Hz. ASTM Type I purified water (with conductivity < 2 nS/cm) was used to prepare all of the aqueous solutions. The quality of water used to rinse the flow system was constantly monitored conductometrically (YSI model 35) and the optical cell was filled with a new sample only when the conductivity of the rinse water decreased to < 1 µS/cm. The UV spectra of the aqueous solutions were obtained using a generic spectrophotometer (OLIS/Cary 14).

Reagents of the highest purity available from Aldrich were used without further purification. For several reagents, purity was of special concern. Most brands of sodium chloride contain bromide and iodide impurity, and ultrapure 99.999% reagent was used. Sodium sulfite is oxidized by dissolved oxygen in water and the solutions were made using deaerated water and used immediately thereafter. *Ferro*cyanide typically contains *ferri*cyanide impurity and its aqueous solutions are not photostable; ultrapure 99.99+% reagent was used under anaerobic conditions and the solutions were stored in the dark. Sodium hydrosulfide was available as a hydrate only (25 wt% of water was assumed from the specifications provided by the manufacturer) and contained traces of colloidal sulfur [35] that reacted with the electron ($k=2\times10^9$ s$^{-1}$ for 1 M HS$^-$). Potassium thiocyanate (99.9+%) contained unidentified impurity that slowly reacted with the electron ($k=4\times10^7$ s$^{-1}$ for 1 M of CNS$^-$). Hydrogen peroxide (1 M standard) was stabilized by traces of tin; no absorbance from the stabilizer was apparent in the UV spectra. High-purity 0.989 N analytical standard KOH was used; the hydrosulfide and hydroxide solutions were purged by dry nitrogen and handled in a nitrogen box.

*Methodology.* The typical QY measurement included determination of four parameters: laser transmission *T* through the sample, the decadic transient absorbance (which was determined at the end of the laser pulse) $\Delta OD_\lambda$ of the photoproduct at wavelength $\lambda$ of the analyzing light, the laser energy $I_{abs}$ absorbed by the sample, and the incident beam energy $I_0$. Several aqueous solutions of the salt were prepared so that the laser transmission through the sample varied between 0.1 and 0.8. Typical kinetics for hydrated electron are shown in Figs. 5 and 3S(a). For each of these solutions, the four parameters were determined and then $\Delta OD_\lambda$ was plotted as a function of $I_{abs}$ (e.g., Figs. 1(a) and 3(a)) and $T$ was plotted as a function of $I_0$ (e.g., Fig. 3(b)) For all systems studied in this work, the latter plots were linear, and the molar extinction coefficients for the anion and the "photoproduct" (hydrated electron and residue X•) can be obtained from these plots using eq. (12) derived in section 3. If the plots of $\Delta OD_\lambda$ vs. $I_{abs}$ were linear (or had a linear initial section), the initial slope of this dependence was determined by a least squares linear regression and QY calculated using eq. (11). This QY was then corrected for the window transparency and noncolinear beam geometry (see above). Some of the plots of $\Delta OD_\lambda$ vs. $I_{abs}$ were very curved (e.g., Fig. 3(a)), even at low laser power. These dependencies were fit using an empirical formula

$$\Delta OD_\lambda \approx A\left[1 - \exp(-B\, I_{abs})\right] \qquad (2)$$

where coefficients $A$ and $B$ were determined by least squares optimization. The initial slope was then estimated from the product $AB$. The use of eq. (2) is justified in section 3. If not stated otherwise, absorption of hydrated electron at 700 nm ($\varepsilon_{700}$=20560 M$^{-1}$ s$^{-1}$ [36]) at the end of the UV pulse (at 30 ns) was used to obtain the electron yield. The anion concentration and light fluence were chosen so that the decay half time of this electron (due to cross recombination) was longer than 1 µs (for most photosystems, this time was 5-10 µs).

**3. Quantum yield measurements: analysis.**

In this section, we derive basic equations needed for the QY determination. Let us introduce $x$, the penetration depth of the excitation light, $L$, the sample thickness, $J(x,t)$, the laser radiance across the unit area of the sample at the delay time $t$ (we assume uniform surface illumination), $J_0(t)$, the radiance of the light incident at the sample at $x=0$, $\beta$ and $\beta_{pr}$, the molar absorptivities of the photolyzed species and the product, respectively, $c(x,t)$, the molar concentration of the substrate in the course of photolysis, $c_0 = c(x, t = -\infty)$, the initial concentration of the photolysate, and $\phi$, the quantum yield of the product. We will assume that (i) the absorption of the laser light by the photoproduct does not change the concentrations of this photoproduct and the substrate and (ii) the photoreactions are complete on a time scale that is much shorter than the duration of the laser pulse. With these assumptions, the absorption of the laser light obeys the following system of partial differential equations

$$\partial J(x,t)/\partial x = -\left(\left[\beta - \beta_{pr}\right] c(x,t) + \beta_{pr} c_0\right) J(x,t) \tag{3}$$

$$\partial c(x,t)/\partial t = -\phi \beta c(x,t) J(x,t) \tag{4}$$

We first consider an ideal situation in which the product does not absorb the excitation light ($\beta_{pr} = 0$). The integration of eq. (3) over $x$ gives

$$J(x,t) = J_0(t) \exp\left(-\beta \int_0^x d\xi \, c(\xi,t)\right) \tag{5}$$

To solve eq. (4), we introduce three new variables: the absorption coefficient $\alpha = \beta c_0$,

$$C(x,t) = c_0^{-1} \int_0^x d\xi \, c(\xi,t) \quad \text{and} \quad q(t) = \phi \beta \int_{-\infty}^t dt \, J(t) \tag{6}$$

With these new variables, eq. (4) is recast as

$$\partial^2 C/\partial x \, \partial q = -\alpha^{-1} \, \partial/\partial x \, \exp(-\alpha C) \tag{7}$$

Integration of this equation yields

$$C(x,q) = x + \alpha^{-1} \ln\left[1 - (1-e^{-q})(1-e^{-\alpha x})\right] \tag{8}$$

Let us introduce the sample-average photoconversion $\Phi = \langle c_{pr} \rangle / c_0$ defined as the ratio of the end-of-pulse, mean-path product concentration

$$\langle c_{pr} \rangle = \frac{1}{L} \int_0^L dx \ (c_0 - c(x, t = \infty)) = c_0 (1 - C(L,Q)/L) \qquad (9)$$

and the initial photolysate concentration $c_0$, where $Q = \phi \beta I_0$ and $I_0 = \int_{-\infty}^{+\infty} dt \ J_0(t)$ is the fluence of the excitation light at $x=0$ (i.e., $Q = q(t = +\infty)$). The mean concentration $\langle c_{pr} \rangle$ of the photoproduct may be determined from the optical density of this photoproduct at the probe wavelength (see below). From eqs. (8) and (9), we immediately obtain $\Phi = -D^{-1} \ln\left[1 - \left(1 - e^{-Q}\right)(1 - T_0)\right]$, where $D = \beta c_0 L$ is the optical density of the sample and $T_0 = e^{-D}$ is the laser transmittance for $Q \to 0$. Note that the conversion $\Phi$ does not depend on the temporal profile $J_0(t)$ of the laser pulse; it depends only on the total photon fluence $I_0$. The total number $N_{abs}$ of photons absorbed by the sample per unit of area is given by

$$N_{abs} = \int_{-\infty}^{+\infty} dt \ [J(0,t) - J(L,t)] = I_0 (1 - T) \qquad (10)$$

where $T$ is the transmittance of the laser pulse. Using eqs. (3) and (4), we obtain

$$\phi = L \langle c_{pr} \rangle / N_{abs} = \left(\Delta OD_\lambda^{pr} / \varepsilon_\lambda^{pr}\right) / I_{abs} \qquad (11)$$

where $\Delta OD_\lambda^{pr}$ is the (decadic) optical density of the photoproduct at wavelength $\lambda$ of the analyzing light (assumed to be collinear with the excitation light), $\varepsilon_\lambda^{pr}$ is the (decadic) molar absorptivity of this photoproduct, and $I_{abs}$ is the total number of absorbed photons per unit area. The latter quantity can be determined by subtracting the transmitted laser power from the incident power. Eq. (11) provides a method for measuring a quantum yield from the initial slope of $\Delta OD_\lambda^{pr}$ vs. $I_{abs}$. Before considering a more realistic case in which the photoproduct also absorbs the excitation light ($\beta_{pr} > 0$), it is useful to derive a simplified expression for the transmittance $T = 1 - D\Phi/Q$ of the laser pulse. For $Q<<1$,

$$\frac{T}{T_0} \approx 1 + \frac{1}{2}(1 - T_0)\left(1 - \frac{\beta_{pr}}{\beta}\right) Q \qquad (12)$$

(Eq. (12) has been generalized for $\beta_{pr} > 0$; to this end, small-$Q$ expansion of eq. (15) was used, as explained below). Formula (12) implies that the initial slope $\partial T/\partial Q\big|_{Q=0}$ of the transmittance plotted as a function of the laser fluence has the same sign as the difference $\beta - \beta_{pr}$: at higher light fluence, the substrate is converted to a product that is less (or more) absorptive and this makes the sample more (or less) transparent to the laser light. Eq. (12) gives a simple recipe for obtaining the extinction coefficients $\beta$ and $\beta_{pr}$ from the transmission data. First, the plot of $T$ vs. $I_0$ is linearized (e.g., Fig. 3(b)), and the extrapolated value at $I_0 \to 0$ gives an estimate for $T_0$. By plotting $D = -\ln(T_0)$ vs. $c_0$, one obtains an estimate for $\beta$. (In practice, we plotted the ratio $OD_{\lambda^*}/L$ of the extrapolated decadic optical density at the excitation wavelength $\lambda^*$ and the optical path $L$ vs. $c_0$ and determined the decadic molar absorptivity $\varepsilon$ of the anion from the slope of the plot; see Figs. 1(b) and 4(a)). This estimate and the quantum yield $\phi$ determined from eq. (11) are used to calculate $Q$ and obtain $T^{-1} \ \partial T/\partial Q\big|_{Q=0}$. The latter quantity is plotted vs. $(1-T_0)/2$ (see, for example, Fig. 4(b)), and the molar absorptivity of the product $\beta_{pr}$ estimated from the slope of this linear plot using eq. (12). Where possible, we have justified this procedure by comparing the molar absorptivities determined from this procedure with the values determined spectrophotometrically.

In the general case, where $\beta_{pr} > 0$, eq. (4) may be rewritten as

$$\partial^2 C / \partial x \partial q = -\partial C / \partial x \ e^{-c_0 \beta_{pr} x} \exp\left[-c_0(\beta - \beta_{pr})C\right] \tag{13}$$

The quantities $\langle c_{pr} \rangle$ and $N_{abs}$ are calculated from $C(x,q)$ using eqs. (9) and (10), where the transmittance $T$ is given by

$$T = e^{-c_0 \beta_{pr} L} \ Q^{-1} \int_0^Q dq \ \exp\left[-c_0(\beta - \beta_{pr}) \ C(L,q)\right] \tag{14}$$

Although in the general case, eq. (13) cannot be solved analytically, the solution can be obtained numerically, using a finite-difference scheme (Fig. 1S(b) and 11S(a) show the typical results of this numerical integration). Since both the product and the substrate absorb laser light, eq. (11) is no longer applicable, and for large $Q$, the quantum yield determined using eq. (11) is smaller than the correct value. Expanding $C(x,q)$ in powers

of $q \ll 1$, we obtain

$$x - C(q,x) \approx \frac{q}{\alpha}(1 - e^{-\alpha x})\left\{1 + \frac{q}{2}\left[\left(1 - \frac{\beta_{pr}}{\beta}\right) - (1 + e^{-\alpha x})\left(1 - \frac{\beta_{pr}}{2\beta}\right)\right]\right\} \quad (15)$$

This expansion was used to obtain eq. (12) and determine the first correction term to the ratio

$$\frac{L\langle c_{pr}\rangle}{\phi N_{abs}} = 1 - \frac{\beta_{pr}}{4\beta}(1 + T_0)\, Q \quad (16)$$

Numerical simulations suggest that for sufficiently small $Q$, the overall dependence of $\Phi$ as a function of $N_{abs}$ can be approximated by an exponential dependence (e.g., Fig. 1S), which justifies the use of empirical eq. (2).

## 4. Results.

This section is organized as follows: First, QY measurements for pure water at 193 nm and hydrogen peroxide at 248 nm are discussed. One-photon ionization of water at 193 nm is facile [24,37] and its efficiency needs to be known to study electron detachment from anions that are poor absorbers of 193 nm light (e.g., chloride and perchlorate). Photodissociation of $H_2O_2$ by 248 nm light is a benchmark system to test the accuracy of our QY measurement (*de facto* providing an actinometric standard) as there have been very reliable QY measurements made by product analysis for this photosystem at 254 nm [38]. Second, we survey those photosystems for which the routine analysis outlined in sections 2 and 3 was insufficient. Third, the data on the anion absorptivities are summarized and compared with the previous spectrophotometric data. Fourth, the discerned "photoproduct" absorptivities estimated by use of eq. (12) are compared with the transient absorption data, where such data exist.

### *4.1. Benchmark photosystems.*

### *4.1.1. Photoionization of water at 193 nm.*

Absorption of 193 nm by water causes its ionization [37]

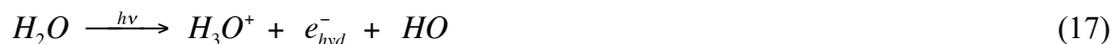

$$H_2O \xrightarrow{h\nu} H_3O^+ + e^-_{hyd} + HO \quad (17)$$

Taking into account the transmittance of the Suprasil windows, a decadic absorption coefficient of 1.4 m$^{-1}$ at 193 nm was estimated, in agreement with Quickenden and Irvin who gave 1.26±0.03 m$^{-1}$ at 196 nm. [39] The absorbance of the electron at 700 nm induced by a 72 mJ/cm$^2$ pulse was ca. 1.75x10$^{-3}$, and the QY was (1.40±0.04)x10$^{-2}$. A QY of (1.27±0.17)x10$^{-2}$ for hydrated electrons was obtained using a d. c. photoconductivity technique for an SF$_6$-saturated aqueous solution, by Bartels and Crowell, which also used an ArF excimer laser [37].

### 4.1.2. Photodissociation of hydrogen peroxide at 248 nm.

Upon UV photoexcitation, hydrogen peroxide dissociates to two hydroxyl radicals [38],

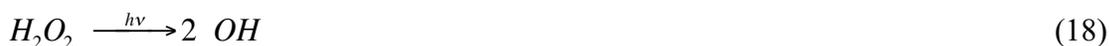

$$H_2O_2 \xrightarrow{h\nu} 2\ OH \quad (18)$$

These hydroxyl radicals can be rapidly reacted with bicarbonate

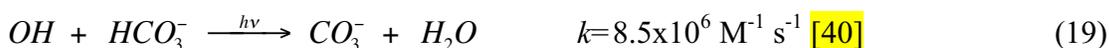

$$OH + HCO_3^- \xrightarrow{h\nu} CO_3^- + H_2O \qquad k=8.5\times10^6\ M^{-1}\ s^{-1}\ [40] \quad (19)$$

to yield carbonate radicals that absorb strongly at 600 nm ($\varepsilon_{600}$=1860 M$^{-1}$ cm$^{-1}$ [41]). We photolysed 24.3 mM and 41.3 mM solutions of H$_2$O$_2$ in the presence of 1 M KHCO$_3$ (*pH* 8.33). Typical kinetics of 600 nm absorbance are shown in Fig. 1S(a). The highest yield of the hydroxyl radicals estimated from our data was 300 µM (Fig. 2S(a)). As explained elsewhere [42], side reactions of hydroxyl radicals (e.g., with peroxide itself [43]) and cross recombination were too slow to compete with rxn. (19) (carbonate radical was formed in 200 ns); the decay of the carbonate radicals was also too slow (5-15 µs) to deplete the concentration of these radicals at *t*=200 ns (Fig. 1S(a)). The plot of the optical density at 600 nm vs. absorbed laser power was slightly curved (Fig. 1(a)) and the plot of the transmission vs. the incident laser power exhibited a negative $\partial T/\partial Q$ slope (Fig. 2S(b)). A quantum yield of 0.443±0.01 for decomposition of H$_2$O$_2$ was obtained from the initial slope of the dependence shown in Fig. 1(b) and a molar extinction coefficient $\varepsilon_{248}$ of 24.8±0.4 M$^{-1}$ cm$^{-1}$ was obtained at 248 nm. Our spectrophotometric measurement gave $\varepsilon_{248}$=26 M$^{-1}$ cm$^{-1}$ vs. the reported 25 M$^{-1}$ cm$^{-1}$ [44] (the absorbance of 248 nm light by bicarbonate was negligible, ca. 0.0206 M$^{-1}$ cm$^{-1}$). Our quantum yield compares well with the primary yield obtained by product analyses: 0.49±0.07 [38] and 0.47±0.03 [45] at 254 nm and 0.45±0.06 at 222 nm [46]. Note that the accuracy of our measurement depends on the accuracy of the extinction coefficient for the carbonate radical, which is ±10% [41].

The negative curvature of the $\Delta OD_{600}$ plot in Fig. 1(a) is accounted for by the absorption of 248 nm light by hydroxyl radicals generated by the laser pulse (as shown elsewhere, these radicals escape from the solvent cage in ca. 30 ps [42]). Czapski and Bielski gave an estimate of $\varepsilon_{248}(OH)=500$ M$^{-1}$ cm$^{-1}$ [47]. Thus, under the conditions of our experiment, up to 20-40% of the 248 nm photons were absorbed by these radicals (Fig. 2S(a), traces (i) and (ii)). Using our estimate for the molar absorptivity of peroxide and eq. (12), $\varepsilon_{248}$ of 508 M$^{-1}$ cm$^{-1}$ was obtained for the hydroxyl radicals from the data of Fig. 2S(b) (this estimate neglects the conversion of OH radicals to $CO_3^-$ radicals that occurred during the 248 nm pulse, which was less than 10%). Using this estimate and solving eq. (13) numerically, we were able to simulate the power dependence shown in Fig. 1(a) (see Fig. 1S(b)). The estimates for the QY and the molar absorptivity of hydrogen peroxide and hydroxyl radicals obtained by our method agree very well with the estimates obtained by completely different techniques.

*4.2. Special cases.*

Most of anion photosystems presented below were much like the benchmark system examined in the previous section: the $\Delta OD_{700}$ plots of hydrated electron absorbance were linear or nearly linear with the absorbed laser power, the extrapolated absorbance at $Q=0$ scaled with the anion concentration, etc. Some of the representative plots for these "perfect" systems are given in the Supporting Information (Figs. 3S to 7S) and the data are summarized in Tables 1 to 3; these results deserve little comment. Other systems exhibited more unusual features, and only those will be discussed below.

*4.2.1. Carbonate (193 nm photoexcitation).*

The peculiarity of the carbonate photosystem is the protic equilibria that involve the carbonate anion. The $pK_a$ values for $H_2CO_3$ are 6.35 and 10.33 (at 25°C), [48] and even dilute solutions of carbonate can be very basic. Let $c$ be the molar concentration of the salt, $K_2$ be the second ionization constant, and $K_w$ be the ionic product for water. For $c>1$ mM, $[OH^-] \approx [HCO_3^-] \approx 2c/\left(1+\left[1+4K_2c/K_w\right]^{1/2}\right)$ and $[OH^-]/[CO_3^{2-}] \approx 0.0146/\sqrt{c}$: the relative hydroxide concentration decreases with the carbonate concentration. Using accurate formulas, one can estimate that for 1.5 and 5.7 mM carbonate solutions (Fig. 8S), the equilibrium concentrations of OH$^-$ and CO$_3^{2-}$ anions are 0.47 mM and 1.02 mM, respectively, and 1 mM and 4.7 mM, respectively. Since at 193 nm hydroxide has 20.6 times higher molar absorptivity than carbonate (as shown below), the presence of hydroxide in the solution has significant implications. In particular, the often cited estimate for molar absorptivity by the carbonate at 193 nm, ca. 630 M$^{-1}$ cm$^{-1}$ (which can be traced back to the 1932 work by Ley and Arends [49]) is incorrect, as it does not take hydroxide into account. As demonstrated below, when the absorption of the UV light by the hydroxide is taken into account, carbonate and bicarbonate have very similar molar absorptivities at 193 nm.

The starting point of this analysis is the determination of $\phi$ and $\varepsilon_{193}$ for hydroxide and bicarbonate at 193 nm. For the latter anion, the protic equilibria are inconsequential since the equilibrium concentration of hydroxide is maintained at ca. 2 µM due to buffering (Fig. 9S). Both of these anions give linear $\Delta OD_{700}$ and $OD_{193}$ plots (Fig. 5S) and the corresponding data are given in Table 1. Fig. 2 demonstrates the concentration dependence for the QY and 193 nm absorptivity for the carbonate system. Both of these quantities systematically increase with the carbonate concentration. Since the QY for electron detachment from bicarbonate is very low (0.07) this photoreaction may be neglected, and the dependencies shown in Fig. 2 were interpreted in terms of competing absorbance of the 193 nm light by CO$_3^{2-}$ and HO$^-$:

$$OD_{193}/L \approx \varepsilon_{OH^-}[OH^-] + \varepsilon_{CO_3^{2-}}[CO_3^{2-}]$$
$$\phi \approx \left\{\varepsilon_{OH^-}[OH^-]\phi_{OH^-} + \varepsilon_{CO_3^{2-}}[CO_3^{2-}]\phi_{CO_3^{2-}}\right\} / (OD_{193}/L) \quad (20)$$

The equilibrium concentrations of hydroxide and carbonate were estimated from the known ionization constants for carbonic acid, [48] so that only two parameters were unknown (the QY and molar absorptivity for the carbonate anion). Those were determined from the least squares fit to eqs. (20) which gave $\varepsilon_{CO_3^{2-}}$ of 150±20 M$^{-1}$ cm$^{-1}$ (vs. 134±2 M$^{-1}$ cm$^{-1}$ for bicarbonate) and a quantum yield $\phi_{CO_3^{2-}}$ of 0.453 (vs. 0.112 for hydroxide). Note that in the above analysis we neglected the absorbance of carbonate radical that is formed in the course of the electron detachment. The latter has molar absorptivity of 685 M$^{-1}$ cm$^{-1}$ at 700 nm which is ca. 3% of that of the electron.

While hydrosulfide anion, HS$^-$, is also a strong base (H$_2$S has $pK_a$'s of 7.05 and 19), [48] the protic equilibria are unimportant since the deprotonation of HS$^-$ is negligible. Even at the lowest hydrosulfide concentration used in this study (0.23 mM) $[HS^-]/[OH^-] \approx 44$. Since the molar extinction coefficients for the hydrosulfide at 193 nm is 1.7 higher than that for the hydroxide, the contribution of the hydroxide to photolysis is negligible.

*4.2.2. Ferrocyanide (248 nm and 193 nm photoexcitation)*

At 193 nm, ferrocyanide exhibits a combination of a near unity quantum yield for electron photodetachment, (Tables 1 and 4) high molar absorptivity of the photolyzed anion [14,23,50] and even higher molar absorptivity for one of the photoproducts (the reported $\varepsilon_{193}$ for ferricyanide is 16300 M$^{-1}$ cm$^{-1}$ [50]). As a consequence, the plots of $\Delta OD_{700}$ vs. $I_{abs}$ are strongly curved (Fig. 3(a)) and photoconversion is extremely efficient (Fig. 10S). The standard procedure used to extract the molar absorptivities using eq. (12) gave 7560±40 M$^{-1}$ cm$^{-1}$ for ferrocyanide (Table 2) and 14300±200 M$^{-1}$ cm$^{-1}$ for the photoproduct (Table 3); see Figs. 3(b) and 4(b). Since the molar absorptivity of hydrated electron at 193 nm is ca. 2300 M$^{-1}$ cm$^{-1}$, the molar absorptivity for ferricyanide is estimated at 12000 M$^{-1}$ cm$^{-1}$ which is considerably lower than 16300 M$^{-1}$ cm$^{-1}$ determined spectrophotometrically [50]. Furthermore, our estimate for ferrocyanide is also lower than the previously published values: 9170 M$^{-1}$ cm$^{-1}$ [14,23] and 8400 M$^{-1}$ cm$^{-1}$ [50]. Another peculiarity is that $\Delta OD_{700}$ vs. $I_{abs}$ plots obtained at different concentrations poorly match though all of these plots exhibit very similar initial slopes (Fig. 3(a), trace (i)). Numerical solution of eq. (13) using the obtained photophysical parameters reproduces the curvature of these plots but gives much less variation with the ferrocyanide concentration (Fig. 11S(a)).

The probable cause of these discrepancies is the considerable ion pairing that occurs in the ferrocyanide system [13,51,52]. In the explored concentration range (< 0.7 mM) only single-cation association is important. Conductivity measurements of Davies [52] suggest an association constant of 176 M$^{-1}$, while the optical measurements of Cohen and Plane [51] suggest a somewhat lower value, 104±8 M$^{-1}$. Using the former value, we estimate that 30% of the anions are paired with a potassium cation in a 0.65 mM ferrocyanide solution. The KFe(CN)$_6^{3-}$ anion is known to be a poorer absorber of the UV light than Fe(CN)$_6^{4-}$ [51]. A closer examination of Fig. 4(a) shows that absorption of the ferrocyanide solution at 193 nm is not exactly proportional to the concentration: there is a slight positive curvature. The latter can be explained assuming the coexistence of free and associated anions that have slightly different molar absorptivities. Using the association constant given by Davies [52], we estimate that free ferrocyanide has $\varepsilon_{193}$ of 7900±140 M$^{-1}$ cm$^{-1}$ whereas the associated anion has 17% lower absorptivity, 6580±400 M$^{-1}$ cm$^{-1}$. Progressive decrease in the absorptivity with the ferrocyanide concentration would account for the concentration dependence of the $\Delta OD_{700}$ vs. $I_{abs}$ plots in Fig. 3(a).

At 248 nm, ferricyanide is a much poorer light absorber than ferrocyanide (865 vs. 4730 M$^{-1}$ cm$^{-1}$; Table 2) and the absorption of laser light causes photobleaching of the solution (Fig. 12S(a)). Both the QY and molar absorptivity of ferrocyanide at 248 nm are significantly lower than these quantities at 193 nm, and the behavior of this anion at 248 nm is much less extreme. Though the plot of $\Delta OD_{700}$ vs. $I_{abs}$ is curved, a single curve can be drawn through the data obtained for different ferrocyanide concentrations (Fig. 12S(c)). Spectrophotometric measurements suggest that the molar absorptivity of KFe(CN)$_6^{3-}$ anion at 280 nm is just 2% lower than that of the Fe(CN)$_6^{4-}$, [51] i.e., the ion pairing appears to be inconsequential for 248 nm photoexcitation. The estimated molar absorptivity of ferrocyanide at 248 nm, ca. 4300 M$^{-1}$ cm$^{-1}$, is close to values given in the literature (4480 M$^{-1}$ cm$^{-1}$ [23], 4730 M$^{-1}$ cm$^{-1}$ [50]). Our estimate for the "product" absorbance is 690±120 M$^{-1}$ cm$^{-1}$ obtained using eq. (12) is probably too low (the hydrated electron itself absorbs 248 nm light with molar extinction coefficient of 600 M$^{-1}$ cm$^{-1}$ [44]).

*4.2.3. Iodide: secondary chemistry.*

In the preceding analysis, it was implied that only hydrated electron absorbs 700 nm analyzing light. For all halides and pseudohalides, with exception of hydroxide, the residue X• undergoes a hemicolligation reaction [53-61]

$$X^{\bullet} + X^{-} \longrightarrow X_2^{-} \qquad (21)$$

(the hydroxide undergoes a reversible deprotonation with the formation of O$^-$ instead, with rate constant of (1.2-1.3)x10$^{10}$ M$^{-1}$ s$^{-1}$; see ref. [12] for more detail). Most of the resulting X$_2^-$ anions absorb in the near-UV and visible (see, for example, ref. [56] for Cl$_2^-$ and ref. [59] for H$_2$S$_2^-$) but only Br$_2^-$ and I$_2^-$ have absorbances at 700 nm. For Br$_2^-$, the molar absorptivity at 700 nm is very low, ca. 380 M$^{-1}$ cm$^{-1}$ (2% of the electron absorbance [36]) and it can be safely neglected. [57,58] For I$_2^-$, this absorbance is not negligible since this anion has a strong absorption band centered at 720-750 nm. [53,54,55,61] According to the survey by Elliot and Sopchyshyn, [61] estimates for the molar absorptivity at the band maximum range from 2120 to 4000 M$^{-1}$ cm$^{-1}$, with 2560 M$^{-1}$ cm$^{-1}$ as the preferred value (which comprises 12.5% of the electron absorbance at 700 nm [36]). For iodine atoms, reaction (21) is rapid ($k_{21}$=(0.88-1.2)x10$^{10}$ M$^{-1}$ s$^{-1}$ [53,54,55]) and some I$_2^-$ formation occurs by the end of the excitation pulse. To assess the degree by which this process interferes with the QY measurement, transient absorption kinetics in N$_2$ and CO$_2$ saturated solutions containing 2 mM iodide were obtained. Carbon dioxide served as an efficient electron scavenger [25] that removed the electron absorbance in ca. 40 ns; the residual absorbance was from I$_2^-$. The progress of electron scavenging can be observed at 590 nm, where I$_2^-$ does not absorb [61] (see Fig. 5, trace (v)). At 700 nm, the kinetics are composite: in the first 50 ns there is rapid decay of the electron absorbance; at later delay times there is the formation of I$_2^-$ which is complete at 200 ns; after this initial growth, the signal slowly decays (Fig. 5, trace (iv)). The formation kinetics of I$_2^-$ can be observed with less interference from the electron at 400 nm, were I$_2^-$ has a strong absorption band: ε=10000 M$^{-1}$ cm$^{-1}$ [61] vs. 2300 M$^{-1}$ cm$^{-1}$ for the electron [44] (Fig. 5, traces (ii) and (iii)). In Fig. 5 we normalized this 400 nm trace (iii) so that the maximum signal at 200 ns matched that at 700 nm (dashed trace). It is seen from this plot that the end-of-pulse 700 nm signal from I$_2^-$ comprises ca. 4 % of the total signal in N$_2$ saturated solution (trace (i) in Fig. 5). Our QY measurements for 193 nm were carried out using dilute iodide solutions (0.1-0.3 mM) and I$_2^-$ formation was negligible. By contrast, the 248 nm measurements were done with more concentrated solutions (0.2-6 mM), and I$_2^-$ formation was important at the higher end of this concentration range. For this reason, only 0.2-1 mM data were used for estimation of the QY; the 1-6 mM data were used for determination of the iodide absorptivity only.

For other photosystems studied in this work (with the exception of the carbonate and bicarbonate), no absorbances in the red from the corresponding radicals X• were found and reactions similar to rxn. (21) are known not to occur for other anions than the halides and pseudohalides. Most of these radicals slowly decay by recombination and disproportionation in the bulk (e.g., refs. [60] and [62]). Given that the anion concentrations were low (typically, < 2 mM), the effect of rxn. (21) on the geminate recombination of $(X^\bullet, e^-_{hyd})$ pairs can be ignored.

### *4.2.4. Perchlorate: the ionic strength effect.*

Perchlorate has very little absorbance at 193 nm, and to observe a signal, very concentrated (2-9 M) solutions were used (Figs. 6(a) and 6(b)). As shown in Part II of this series, [21] for all anions, the addition of (chemically inert) salts, such as $Na_2SO_4$ and $NaClO_4$, causes a large decrease in the QY for electron photodetachment, ca. 6-10% per 1 M of ionic strength, and this pertains to the sulfate and perchlorate anions themselves. While sulfate is a relatively strong absorber at 193 nm (46±7 $M^{-1}$ $cm^{-1}$, Table 1), so that the QY can be determined in dilute solutions (4-20 mM) for which the ionic strength effect is minor, the extrapolated molar absorptivity for perchlorate (see below) is very low, ca. 0.57 $M^{-1}$ $cm^{-1}$, and the use of concentrated solutions is unavoidable. In such solutions, two other effects occur: the CTTS absorption band shifts to the blue [1] and the absorption band of the hydrated electron itself shifts to the blue (resulting in a 20% loss of 700 nm absorbance in 9 M $NaClO_4$ solution [21,63]). Due to the anion band shift, the plot of $OD_{193}$ vs. [$ClO_4^-$] is nonlinear (Fig. 6(b)), and the molar absorptivity given above was obtained from a polynomial fit to this plot and its extrapolation to zero concentration of the perchlorate. The plots of $\Delta OD_{700}$ vs. $I_{abs}$ exhibit a negative curvature, suggesting a competition between one- and two-photon excitation (Fig. 6(a)). The plots were fit using the formula $\Delta OD_{700} \approx A\ I_{abs} + B\ I^2_{abs}$. Both of these coefficients linearly decrease with perchlorate concentration with a slope of 11% per 1 M of ionic strength (Fig. 6(b)). Similar slopes of 6-10% per M were observed for other anions in concentrated sulfate and perchlorate solutions (such as hydroxide, iodide, bromide, and sulfite). [21] The extrapolated quantum yield for electron photodetachment from the perchlorate (at infinite dilution) is ca. $4 \times 10^{-3}$, which is 3 times lower than the QY for the water itself. In 9 M perchlorate solution, almost no one-photon electron detachment occurs; the electron generation is completely biphotonic (Fig. 6(a)).

For other anions, this biphotonic excitation was negligible. Furthermore, for all systems other than perchlorate, sufficiently high concentration of the anions can be used to avoid the photoexcitation of water at 193 nm (section 4.1.1). A more serious concern is the possibility that an impurity rather than the anion being studied yields electrons upon photoexcitation. Such a dilemma presents itself for all anions that exhibit low (<0.02) quantum yield for electron photodetachment. The presence of impurities in thiocyanate, chlorate, and hydrosulfide solutions can be discerned from the fact that electron half time was considerably shortened in these solutions (section 2). We believe that with the possible exception of thiocyanate at 248 nm and perchlorate at 193 nm, the involvement of impurity is unlikely because molar absorptivities of the low-QY anions (such as hydrosulfide, nitrite, nitrate and chlorate; see Table 1) were actually quite large; moreover, these absorptivities compared well with the spectrophotometric data (Table 2). None of the impurity ions specified by the manufacturer (that are, typically, poor light absorbers like sulfate, chloride, and transition metal cations) can account for the observed QYs at their expected concentrations.

*4.3. Molar absorptivities for aqueous anions.*

Table 2 provides a comparison between the molar absorptivities were obtained in this study and those obtained spectrophotometrically, where such data are available. It must be stressed that for some of these anions, conflicting estimates of the absorptivities have been reported, especially at 193 nm, and it is not clear which values should be compared to ours. For most of these anions, at least one of the reported values was close to our estimate. For some anions (ferrocyanide, carbonate) there was a significant difference; the origin of these discrepancies has been addressed in section 4.2. A surprisingly large scatter exists in the literature for iodide at 248 nm (which is, after all, the benchmark system for CTTS studies): 400 to 900 $M^{-1}$ $cm^{-1}$ (Table 2). We have carried out our own spectrophotometric measurement using 0.7 mM iodide solution and obtained 870±10 M$^{-1}$ cm$^{-1}$ - in perfect agreement with the data of Tables 1 and 2.

*4.4. Estimates for "photoproduct" absorptivities.*

Estimates for the molar absorptivity of "photoproducts" obtained using eq. (12) are given in Table 3. Not all of the systems studied in this work provided good quality data for such a measurement, though the sign of $\partial T/\partial Q|_{Q=0}$ (and, therefore, the sign of $\beta - \beta_{pr}$) can be determined for all of these photosystems (these signs are given in Table 1). The analysis of these signs suggests that at 193 nm (248 nm) $\beta$ was only greater than $\beta_{pr}$ for anions whose absorptivities were greater than 3000 M$^{-1}$ cm$^{-1}$ (400 M$^{-1}$ cm$^{-1}$, respectively). These values are not surprising since one of the "photoproducts", hydrated electron, has an estimated molar absorptivity of 2300 M$^{-1}$ cm$^{-1}$ at 193 nm and 600 M$^{-1}$ cm$^{-1}$ at 248 nm [44]. To result in positive $\partial T/\partial Q|_{Q=0}$ slope, the absorptivity of the anion should be at least greater than that of the hydrated electron. For hydroxide excitation at 193 nm, this slope is only very slightly positive. Given the hydroxyl absorptivity of 500 M$^{-1}$ cm$^{-1}$ at this wavelength, [47] one obtains that hydroxyl and electron in sum absorb 193 nm light with ε≈2800 M$^{-1}$ cm$^{-1}$ - which is fairly close to the discerned "critical" value of 3000 M$^{-1}$ cm$^{-1}$.

For sulfite radical, Hayon et al. [62] give an estimate of ε$_{280}$=630 M$^{-1}$ cm$^{-1}$, from which ε$_{248}$≈1390 M$^{-1}$ cm$^{-1}$ can be obtained using the spectrum of this radical given in the same work. Thus, at 248 nm $\varepsilon(SO_3^-) + \varepsilon(e_{hyd}^-)$≈1990 M$^{-1}$ cm$^{-1}$ vs. 1850±190 M$^{-1}$ cm$^{-1}$ obtained from our data on electron detachment from sulfite (Table 3). For thiosulfate radical, Devonshire and Weiss [58] give ε$_{375}$≈1720 M$^{-1}$ cm$^{-1}$, from which an estimate of ε$_{248}$≈5650 M$^{-1}$ cm$^{-1}$ can be obtained using the spectrum of this radical given in the same work. Adding in the electron absorbance, one obtains 6250 M$^{-1}$ cm$^{-1}$ vs. 5046±280 M$^{-1}$ cm$^{-1}$ in Table 3.

The extinction coefficients for halide atoms and sulfate radical at 193 nm are not known. From general considerations, one can expect that the absorptivity of SO$_4^-$, Cl, Br, and I increases with the polarizability. This trend is reflected in the data of Table 3. For iodide, the "photoproduct" (iodine atom and hydrated electron) has molar absorptivity that is much greater than 2300 M$^{-1}$ cm$^{-1}$ (given by Nielsen et al. for the molar extinction coefficient of hydrated electron [44]), i.e., it is certain that the iodine atom absorbs 193 nm light. For bromide, the molar absorptivity of the "photoproduct" is close to that of the electron. For the other two anions, $SO_4^{2-}$ and $Cl^-$, our estimates seem to be unrealistically low. This is not too surprising, since for these two anions even a relatively small error in the ratio $\beta_{pr}/\beta$ estimated using eq. (12) would cause large variation in the estimated "photoproduct" absorptivity (these ratii are, respectively, 21.2 and 6.1). While these latter estimates may not be too reliable, it is clear that in many CTTS systems the absorption of the excitation light by the hydrated electron and the radical residue of the parent anion routinely occurs in the course of photoexcitation. A pertinent question is, can this excitation change the QY? It is known from the kinetic studies of water ionization by Barbara and coworkers [64] that UV excitation of hydrated electron changes the escape yield of this electron due to the occurrence of geminate recombination suppression: the electron is excited into the conduction band and thermalizes further from the parent hole, so that the resulting solvated electron has higher probability of escape. A similar decrease in the free electron yield after electron excitation has been observed for solvated electrons generated in sodide CTTS in tetrahydrofuran. [8] According to Schwartz and coworkers, [8,9,10] only electrons that are close to the residual sodium atom (and exhibit fast exponential kinetics) can be depleted via this photostimulated recombination. If that is also the case for aqueous CTTS photosystems, the UV excitation of electrons generated within the laser pulse is of little concern if this pulse is much longer than 10-20 ps, as is the case in this study.

**5. Discussion**

*5.1. Comparison with the previous estimates.*

In Table 4, our estimates for the quantum efficiency of free hydrated electron generation in electron photodetachment from several anions (Table 1) are compared with (i) flash photolysis data of Iwata et al. for 193 nm, 222 nm, and 248 nm photoexcitation of the same anions, [24] (ii) ultrafast pump-probe kinetic and photometry data of Lian et al. for 200 nm photoexcitation (both the prompt and extrapolated free electron yields are given) [17], and (iii) the estimates obtained using product analysis (mainly, using nitrogen evolution in rxn. (1) for quantifying the electron yield) for 185, 229, and 254 nm (Hg line) photolysis. All in all, the quantitative agreement with the data of Iwata et al. [24] and the product analyses of Dainton and Fowles [29] is quite reasonable. Reassuring as that may seem, this good correspondence is a result of error compensation (e.g., Iwata et al. [24] used a 10% lower estimate for the molar absorptivity of hydrated electron). It appears that Iwata et al. [24] systematically underestimated the QYs due to the occurrence of light absorbance by photoproduct(s). The good agreement between our 193 nm data for $Br^-$ and $HO^-$ and 185 nm data of Dainton and Fowles [29] is reassuring because it is known from the picosecond kinetic studies of Lian et al. [17] that for these two anions, most of the geminate decay occurs in less than 50 ps and, therefore, geminate scavenging via rxn. (1) is minimal. This is not the case for polyvalent anions such as sulfate, and QYs estimated from the $N_2$ evolution for such anions can be overestimated considerably. For ferrocyanide, almost no geminate decay was observed in the first 500 fs after 200 nm photoexcitation of this anion, [17] and the closeness of the QY in 193 nm photoexcitation to the prompt (near unity) electron yield is not surprising. For 241 nm photoexcitation of ferrocyanide, Lenchenkov et al. [13,14] observed ca. 20% decrease in the electron concentration in the first 1.5 ns after the laser pulse; i.e., the geminate recombination is efficient and the QYs determined using the product analysis (in this case, $N_2$ and ferricyanide yields) depended strongly on the method used (Table 4).

For 254 nm photolysis of halide anions, especially iodide, the agreement between our QY estimates and those obtained by the product analysis depends on which data sets are used (Table 4). As explained in section 1.1, the scatter in the latter estimates is traceable to corrections that were made to take into account scavenging of geminate electron in reaction (1). Most of the workers made no provision for such a reaction, and their results are suspect. Jortner et al. [16] made the most consistent effort to make such a correction (without the benefit of knowing the geminate decay kinetics) and their best estimate for the QY (0.28-0.29) is close to ours. However, it should be stressed that some of their estimates were obtained by scavenging of the electron by sulfuric acid (with or without N₂O in the solution) and determining the electron yield from $H_2$ evolution; these particular sets of data are certainly incorrect since the QY for electron photodetachment depends on the ionic strength [21] (this effect partially explains the inconsistencies noted by Bradforth and coworkers [4]).

*5.2. Wavelength dependence.*

For all aqueous anions studied in this work that yield hydrated electrons in the course of 248 nm photoexcitation, the QY of electron detachment in 193 nm photoexcitation is higher than in the 248 nm photoexcitation. When the QY data of Table 1 are complemented by 200 nm and 225 nm data obtained by Lian et al. [17] and Iwata et al., [24] respectively (see Table 4), the general trend of the increase of the free electron quantum yield with the excitation energy becomes apparent. As mentioned in the Introduction, similar increase in the electron yield has been observed by Shirom and Stein [23] for photoexcitation of ferrocyanide and by Bradforth and coworkers for one- and two-photon excitation of iodide [11].

It is tempting to explain this behavior, in all of these cases, by an increase in the efficiency of direct ionization that is known to be more efficient at higher excitation energy [11]. However, for some polyatomic anions, electron photodetachment is not the only photoreaction. From their kinetic studies, Lenchenkov et al. [14] concluded that electron photodetachment from ferrocyanide by $\lambda > 224$ nm light (whose absorbance mainly facilitates metal-to-ligand transitions) is preceded by ultrafast internal conversion from the corresponding excited state to a dissociative CTTS state. It is not obvious that such a conversion is 100% efficient for all wavelengths. At lower excitation energy, $\lambda > 250$ nm, ferrocyanide undergoes photoaquation ($CN^-/H_2O$ ligand exchange) [65] and thiocyanate photodissociates to sulfur and cyanide (for $\lambda > 236$ nm). [60] These photoreactions may occur, to a lesser degree, at higher excitation energies and compete with the electron detachment. In such a case, the increase in the electron yield with the photoexcitation energy may reflect a competition of the CTTS state dissociation with these side photoreactions rather than a competition between the latter and the direct ionization. Furthermore, for some of these anions the electron detachment could be concerted with other reactions. For example, photoexcitation of bicarbonate is likely to involve a concerted electron *and* proton transfer because the HCO$_3$ radical is extremely unstable in water. [40] This might account for a large difference between the QYs for electron detachment in 193 nm photoexcitation of bicarbonate and carbonate. Electron photodetachment from perchlorate is another possible example of electron transfer concerted with a rapid reaction of the residual radical; in this case, water oxidation (the ClO$_4$ radicals were observed by IR spectroscopy in a low temperature neon matrix only [66]).

For several of the anions studied in this work (nitrate, nitrite, carbonate, bicarbonate, chlorate, and perchlorate) it is not settled whether their absorption bands in the UV actually involve a CTTS state, as their intramolecular transitions are known to occur in the same spectral region. [1] The occurrence of hydrated electrons does not settle the issue since these electrons can be generated via a direct ionization that does not involve a CTTS state. Nevertheless, relatively large QYs for bicarbonate, nitrite, and, especially, carbonate suggest that their CTTS states are indeed involved. For nitrite, this involvement has been suggested by Blandamer and Fox [1] who deduced it from the polarity effects on the position of high-energy, high-intensity subband.

*5.3. Prompt vs. free electron yield.*

The QY for the free electron is given by a product of two quantities: the prompt QY of the electron and the fraction of photogenerated electrons that escape geminate recombination. Only ultrafast kinetic studies can give an estimate for these two quantities separately, and more detailed discussion of the prompt quantum yields will be given elsewhere. Combining the 220-250 nm results of Bradforth and coworkers [3-7,11] and 200 nm results of Lian et al. [17] (Table 4), it appears that the prompt QY for halide anions is near unity across the entire CTTS band whereas the prompt QY for pseudohalides is much lower, ca. 0.3-0.4 at the band maximum. The constancy of the prompt QY across the CTTS band is not a general property of the aqueous anions. For thiocyanate, the fraction of the escaped electrons increases by ca. 20% from 225 nm to 200 nm. [17] It is unlikely that this fraction changes by an order of magnitude between 248 nm and 225 nm, as would be required to account for the QY of ca. 0.02 (at 248 nm, Table 1) provided that the prompt QY is constant between 200 and 248 nm (Table 4). The free electron QY for thiosulfate and hydrosulfide photoexcited by 248 nm light is > 10 times less than that for their 193 nm photoexcitation; again, it seems unlikely that a change in the escape fraction can be sufficiently great to account for this large decrease in the free electron yield.

**6. Conclusion.**

Time resolved transient absorption spectroscopy has been used to determine quantum yields for electron detachment in 193 and 248 nm laser photolysis of fifteen aqueous anions, including several anions for which no QYs have been reported (Table 1). Molar extinction coefficients for these anions at the laser wavelength were also determined (Tables 1 and 2) so that accurate cross sections for electron photodetachment can be calculated from the data of Table 1. Furthermore, we have estimated the molar absorptivity of several photoproducts at these wavelengths; those estimates are given in Table 3. Our results for 193 nm photolysis of halide and pseudohalide anions correspond well with the previous estimates by Iwata et al. [24] and Dainton and Fowles, [29] and we suggest using these photosystems as convenient short-wave actinometric standards for aqueous photochemistry. We also confirm the previous measurements of the quantum yields of water ionization (193 nm one photon excitation) [37] and hydrogen peroxide dissociation (248 nm photoexcitation); [38,45] the latter photosystem also provides a convenient actinometric standard for time-resolved studies.

It is shown that the QY for free electron formation systematically increases with the excitation energy, due to the increased efficiency of direct ionization at higher excitation energy and blocking of the alternative photoreaction routes. Relatively large QYs for electron generation in 193 nm photolysis of bicarbonate, nitrite, and carbonate suggest that their CTTS states are involved in the photoreaction, as was suggested by Blandamer and Fox. [1] Our data indicate that for polyatomic anions, both the free electron QY and the prompt electron QY increase with the photoexcitation energy, whereas for ferrocyanide and halides, the prompt electron QY is near unity across the whole CTTS band(s). Some of the anions were shown to be involved in thermal reactions (protic equilibria and ion pairing) that have significant implications for their photochemistry. Furthermore, as shown in Part II of this series [21] (and, for perchlorate, in this paper), the QY of electron detachment and the fraction of escaped electrons both decrease with an increase in the ionic strength when the latter is in the molar range. All of these observations point to the complexity of the primary photoprocess; there is little support for a popular claim that CTTS systems provide "simple" models for studying more involved electron transfer reactions.

**7. Acknowledgement.**

We thank Prof. S. E. Bradforth of USC for suggesting the subject of this study and sharing his unpublished results. IAS thanks Drs. D. M. Bartels of NDRL and S. V. Lymar of BNL for many useful discussions. The research at the ANL was supported by the Office of Science, Division of Chemical Sciences, US-DOE under contract number W-31-109-ENG-38.

***Supporting Information Available:*** (1.) Captions to Figs. 1S to 12S; (2.) A PDF file containing Figs. 1S to 12S. This material is available free of charge via the Internet at http://pubs.acs.org.

**Table 1.**

Decadic molar absorptivities (ε) of selected aqueous anions and absolute quantum yields (QY) for generation of free hydrated electrons from these inorganic anions by single 193 nm and 248 nm photon excitation (dilute aqueous solutions at 25°C). [a]

| aqueous anion | 248 nm (5 eV) | | 193 nm (6.43 eV) | |
|---|---|---|---|---|
| | QY | ε, $M^{-1}$ $cm^{-1}$ | QY | ε, $M^{-1}$ $cm^{-1}$ |
| $[Fe(CN)_6]^{4-}$ | 0.674±0.009 | 4273±17 (+) | 1.018±0.050 | 7560±43 (−) |
| $SO_3^{2-}$ | 0.108±0.001 | 49±1 (−) | 0.391±0.011 | 5990±84 (+) |
| $SO_4^{2-}$ | - | - | 0.833±0.023 | 46±7 (−) |
| $S_2O_3^{2-}$ | 0.0252±0.0003 | 412±2 (−) | 0.518±0.016 | 3104±36 (−) |
| $CO_3^{2-}$ | - | - | 0.453±0.006 | 150±20 |
| $HCO_3^-$ | - | - | 0.070±0.005 | 134±2 |
| $Cl^-$ | - | - | 0.463±0.034 | 320±12 (−) |
| $Br^-$ | - | - | 0.365±0.011 | 11502±154 (+) |
| $I^-$ | 0.286±0.008 | 885±3 (+) | 0.497±0.018 | 13250±425 (+) |
| $OH^-$ | - | - | 0.112±0.0015 | 3099±15 (+) |
| $CNS^-$ | 0.0186±0.005 | 60±3 (−) | 0.306±0.003 | 10450±3100 (+) |
| $HS^-$ | 0.0142±0.0009 | 2170±30 (0) | 0.298±0.015 | 5342±73 (+) |
| $NO_3^-$ | - | - | 0.0064 | 7030 |
| $NO_2^-$ | - | - | 0.066±0.002 | 3517±32 (+) |
| $ClO_3^-$ | - | - | 0.012±0.001 | 424±11 (−) |
| $ClO_4^-$ | - | - | 0.0043±0.0001 | 0.565±0.007 |

a) the signs of $\partial T/\partial Q|_{Q=0}$ (see eq. (12)) are given in the parentheses; the error limits indicate the standard deviation of least squares fits, as explained in the text.

**Table 2.**

Literature data for decadic molar extinction coefficients ε of aqueous anions (in $M^{-1}$ $cm^{-1}$) at 193 and 248 nm compared with the values obtained in this work (Table 1).

| anion | 248 nm (5 eV) literature | this work | 193 nm (6.43 eV) literature | this work |
|---|---|---|---|---|
| $[Fe(CN)_6]^{4-}$ | 4480 [23], 4730 [50], 6000 [67] | 4273 | 9170, [a] 8400 [50] | 7560 |
| $SO_3^{2-}$ | 30 [68], 46 [62] | 49 | - | 5990 |
| $SO_4^{2-}$ | - | - | 125 [68] | 46±7 |
| $S_2O_3^{2-}$ | 360 [69] | 412 | - | 3104 |
| $CO_3^{2-}$ | - | - | 630 [49], 200 [50] | 150 |
| $HCO_3^-$ | - | - | 160 [49] | 134 |
| $Cl^-$ | - | - | 800 [68], 470 [50] | 320 |
| $Br^-$ | - | - | 10000 [68], 12000 [50,70] | 11502 |
| $I^-$ [e] | 400 [68], 676 [69], 900 [50] | 885 | 14000 [50,70] | 13250 |
| $OH^-$ | - | - | 4000 [79], 3000 [50] | 3099 |
| $CNS^-$ | <100 [68], 60-70 [50] | 60 | 22000, [b] 12600 [50] | 10450 |
| $HS^-$ | 1000 [68] | 2170 | 10000, [c] 6300 [50] | 5342 |
| $NO_3^-$ | - | - | 8000 [50] | 7030 |
| $NO_2^-$ | - | - | 3550 [50] | 3517 |
| $ClO_4^-$ | - | - | < 1 [78] | 0.565 |

a) calculated from the data of refs. [14] and [23]; b) calculated from the data of refs. [72] and [73]; c) calculated from the data of refs. [68] and [71].

**Table 3**

Estimated decadic molar absorptivity (in $M^{-1}$ $cm^{-1}$) for photoproduct(s) of 193 and 248 nm laser excitation of selected aqueous anions (obtained using eq. (12)).

| anion | 248 nm | 193 nm |
|---|---|---|
| $[Fe(CN)_6]^{4-}$ | 690±120 | 14320±210 |
| $S_2O_3^{2-}$ | 5046±280 | 6640 |
| $SO_3^{2-}$ | 1850±190 | - |
| $SO_4^{2-}$ | - | ≈980 |
| $I^-$ | - | 5565±130 |
| $Br^-$ | - | 2280±310 |
| $Cl^-$ | - | ≈1930 |

**Table 4.**

Absolute quantum yields of free hydrated electrons for photoexcitation of inorganic anions by 193 and 248 nm light (25$^{o}$C aqueous solution). [a]

| aqueous anion | this work | | ref. [24] | | | Ref. [17], 200 nm | | product analyses (limiting yields) [a] |
|---|---|---|---|---|---|---|---|---|
| | 248 nm | 193 nm | 248 nm | 222 nm | 193 nm | $t=0$ | $t=\infty$ | |
| $[Fe(CN)_6]^{4-}$ | 0.67 | 1.02 | - | - | - | 1.0 | 0.96 | 0.66 (254) [15] 0.65 (254) [23] 0.45 (254) [27] 0.66, 0.46 (254) [74] 0.88 (214) [23] 0.55 (254) [23] 0.35 (254) [75] |
| $SO_3^{2-}$ | 0.11 | 0.39 | | - | | 0.35 | 0.23 | - |
| $SO_4^{2-}$ | - | 0.83 | - | - | 0.73 | - | - | 0.71 (185) [29] |
| $Cl^-$ | - | 0.46 | - | - | 0.41 | - | - | 0.43 (185) [29] 0.49 (185) [b] 0.99 (185) [30] |
| $Br^-$ | - | 0.37 | - | - | 0.35 | 0.91 | 0.29 | 0.21 (229) [30] 0.34 (185) [29] 0.36 (185) [b] 0.68 (185) [30] |
| $I^-$ | 0.29 | 0.50 | 0.22 | 0.27 | 0.47 | 0.90 | 0.42 | 0.22 (254) [26] 0.29 (254) [16] 0.24 (229) [30] 0.97 (185) [30] |
| $OH^-$ | - | 0.11 | - | - | 0.11 | 0.38 | 0.09 | 0.11 (185) [29,76] |
| $CNS^-$ | 0.02 | 0.30 | 0.015 | 0.019 | 0.27 | 0.30 | 0.28 | - |

a) photolysis wavelengths (nm) are given in the parenthesis; b) calculated in ref. [29] from the results of ref. [30].

**Figure captions.**

**Fig. 1.**

(a) Maximum transient absorbance of 600 nm analysing light in 248 nm laser excitation of a solution containing 24.3 mM *(open squares)* and 41.3 mM *(open circles)* hydrogen peroxide and 1 M KHCO$_3$ plotted vs. the absorbed laser power. The kinetic traces are shown in Fig. 1S(a). This 600 nm absorbance signal is from carbonate radical formed by rxn. (19); it is proportional to the yield of hydroxyl radicals photogenerated in rxn. (18). The solid line is a least squares exponential fit using eq. (2) (compare this plot with the simulated dependence given in Fig. 1S). The negative curvature is due to the absorbance of 248 nm excitation light by photogenerated OH radicals (see Fig. 2S and section 4.1.2). (b) The concentration dependence of $OD_{248}/L$ ratio (hereafter $L$ is the optical path of the excitation light). The optical density $OD_{248}$ of the solution was determined by extrapolation of the transmission data of Fig. 2S(b) to zero $Q$. Molar absorptivity of hydrogen peroxide at 248 nm can be determined from the slope of this plot.

**Fig. 2.**

The [Na$_2$CO$_3$] dependence of the optical density of the carbonate solution at the laser wavelength *(to the right)* and "quantum yield" of photoelectron *(to the left)* in 193 nm laser photoexcitation of this solution. The concentration dependence of the "quantum yield" $\phi$ (given by eq. (20) and nonlinearity of $OD_{193}$ vs. [carbonate] plot originate through the photoexcitation of hydroxide which is present in the carbonate solution due to (dark) protic equilibria. The solid lines were obtained by least squares fit of the data using eq. (20).

**Fig. 3.**

(a) End-of-pulse transient absorbance of photoelectron (observed at 700 nm) in 193 nm photoexcitation of aqueous potassium ferrocyanide plotted vs. the absorbed laser power. The concentration of ferrocyanide was 50 µM *(open squares),* 100 µM *(open triangles),* 240 µM *(open diamonds),* 480 µM *(open circles),* and 695 µM *(filled circles).* The solid lines are least squares fits to eq. (2). Straight line (i) indicates the initial slope of these power dependencies; this slope does not depend on the ferrocyanide concentration. (b) Transmission of the 193 nm laser light in the same photosystem as a function of dimensionless parameter $Q = \phi \beta I_0$ (same symbols are used as in (a)). The negative slope indicates that $\beta_{pr} > \beta$ (eq. (12)). See Fig. 4(b) for the plot of $1/T_0 \left(\partial T/\partial Q\right)_{Q \to 0}$ vs. $(1-T_0)/2$. Fig. 11S exhibits a simulation of these plots using the formulas derived in section 3.

**Fig. 4.**

(a) The concentration dependence of path-scaled optical density $OD_{193}$ *(open circles)* of ferrocyanide solution at the photoexcitation wavelength (see Fig. 3). This density was determined by extrapolation of transmission $T$ of 193 nm laser light to zero power (i.e., $Q=0$) by linearization of the plots shown in Fig. 3(b). The solid line is a least squares line drawn through the experimental points; the residuals are given *vide supra*. (b) The plot of $-1/T_0 \left(\partial T/\partial Q\right)_{Q \to 0}$ vs. $(1-T_0)/2$ for five concentrations of ferrocyanide given in Fig. 3(b). A linear dependence is given by eq. (12); the ratio $\beta_{pr}/\beta$ can be determined from the slope of this plot.

**Fig. 5.**

Formation and decay kinetics of transient absorbance in 248 nm laser photolysis of 2 mM sodium iodide solution (ca. 0.03 J/cm$^2$ flux of incident photons per pulse). Traces (i) and (ii) were obtained from N$_2$-saturated solution; traces (iii), (iv), and (v) were obtained from CO$_2$-saturated solution. Carbon dioxide served as a hydrated electron scavenger. Traces (i) and (iv) are of 700 nm absorbance, traces (ii) and (iii) are of 400 nm absorbance, and trace (v) is of 590 nm absorbance. The dashed line is scaled down trace (iii). The delayed formation of transient absorbance in traces (ii), (iii), and (iv) is due to the generation of I$_2^-$ in rxn. (21). This reaction is faster in more concentrated iodide solutions (see Fig. 3S). See section 4.2.3 for more detail.

**Fig. 6.**

(a) Transient absorbance of 700 nm light at the end of excitation pulse for 193 nm laser photolysis of (i) 9 M, (ii) 4.5 M, and (iii) 2.08 M NaClO$_4$ vs. the absorbed laser power. All three power dependencies can be fit by a formula $\Delta OD_{700} \approx A\, I_{abs} + B\, I_{abs}^2$ (in which the first and the second term correspond to mono- and bi- photonic excitation of the photosystem, respectively). (b) *To the right:* The concentration plot of the optimum coefficient $A$ thus obtained (which is proportional to $\phi$) vs. perchlorate concentration. The QY linearly decreases with [ClO$_4^-$], the extrapolation of this plot to infinite dilution gives the photoelectron yield given in Table 1. *To the left:* The concentration plot of the $OD_{193}/L$ ratio for perchlorate. The observed nonlinearity is caused by a systematic blue shift of the absorption band of perchlorate with the ionic strength of the solution (see ref. [21] for more detail).

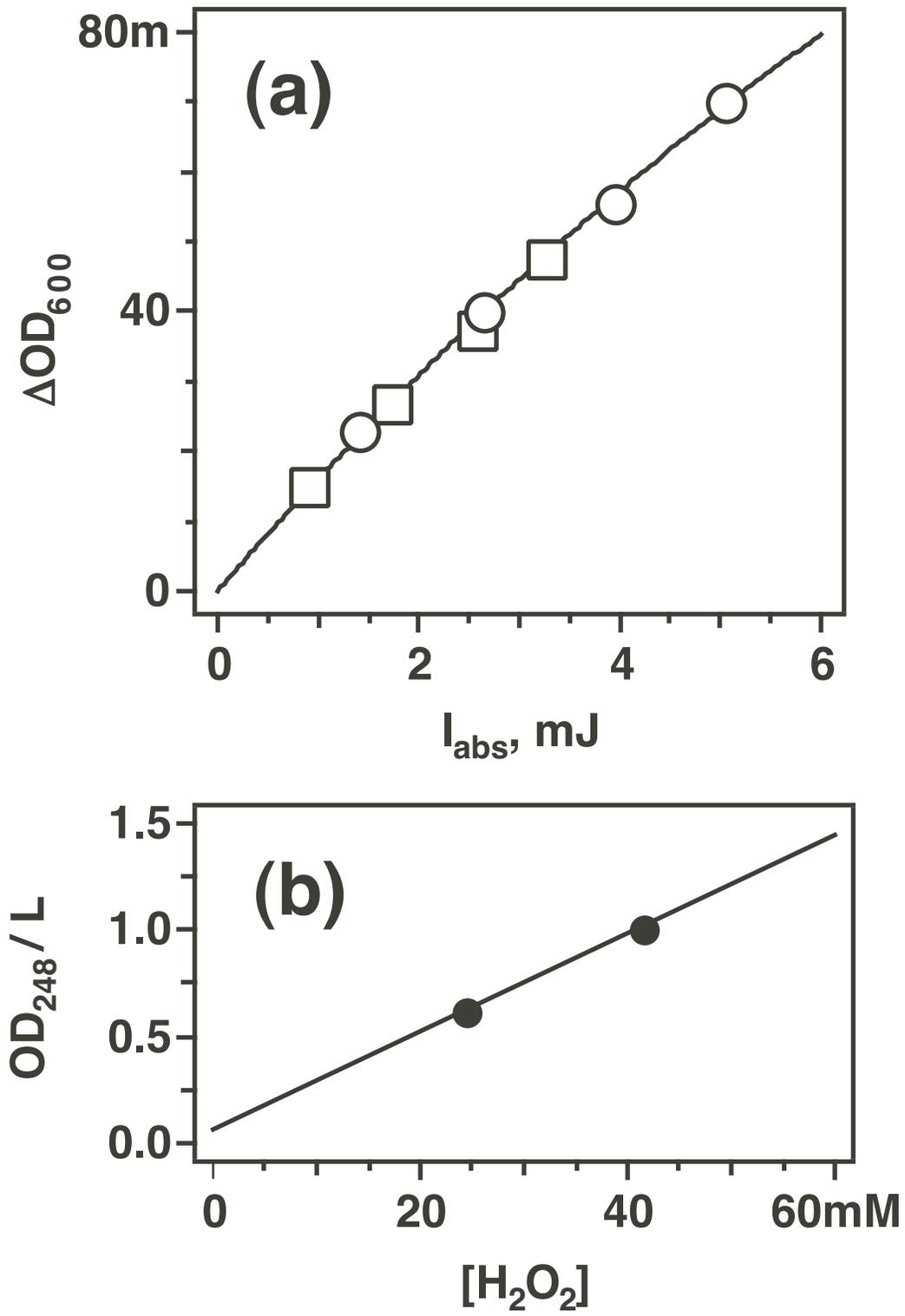

**Figure 1; Sauer et al.**

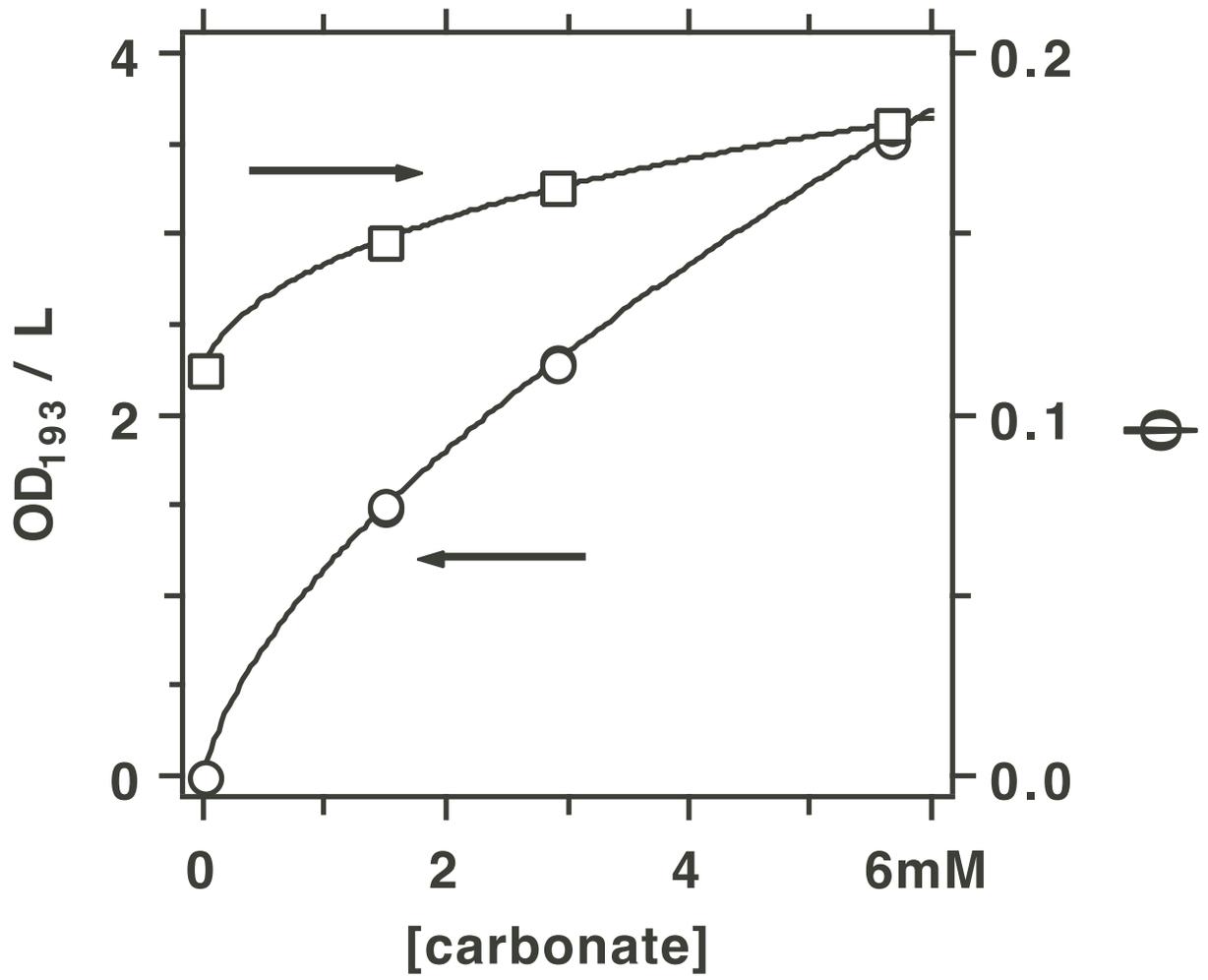

**Figure 2; Sauer et al.**

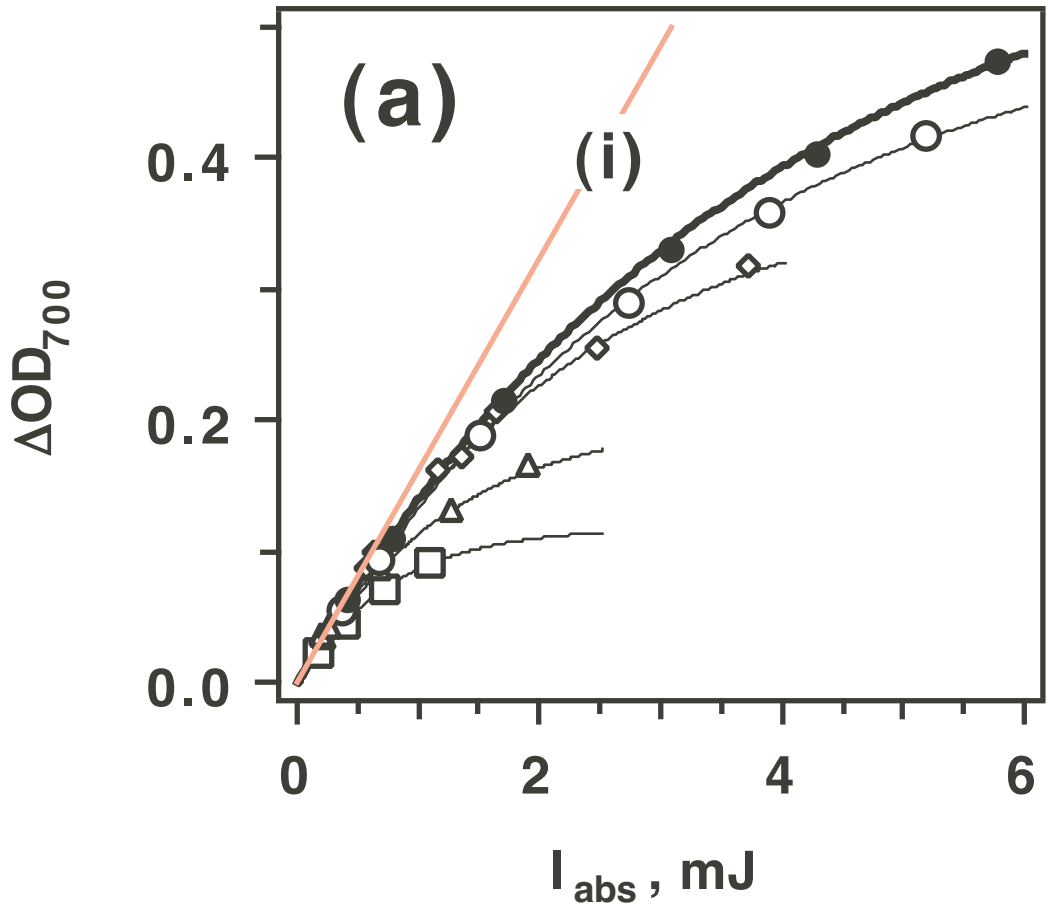
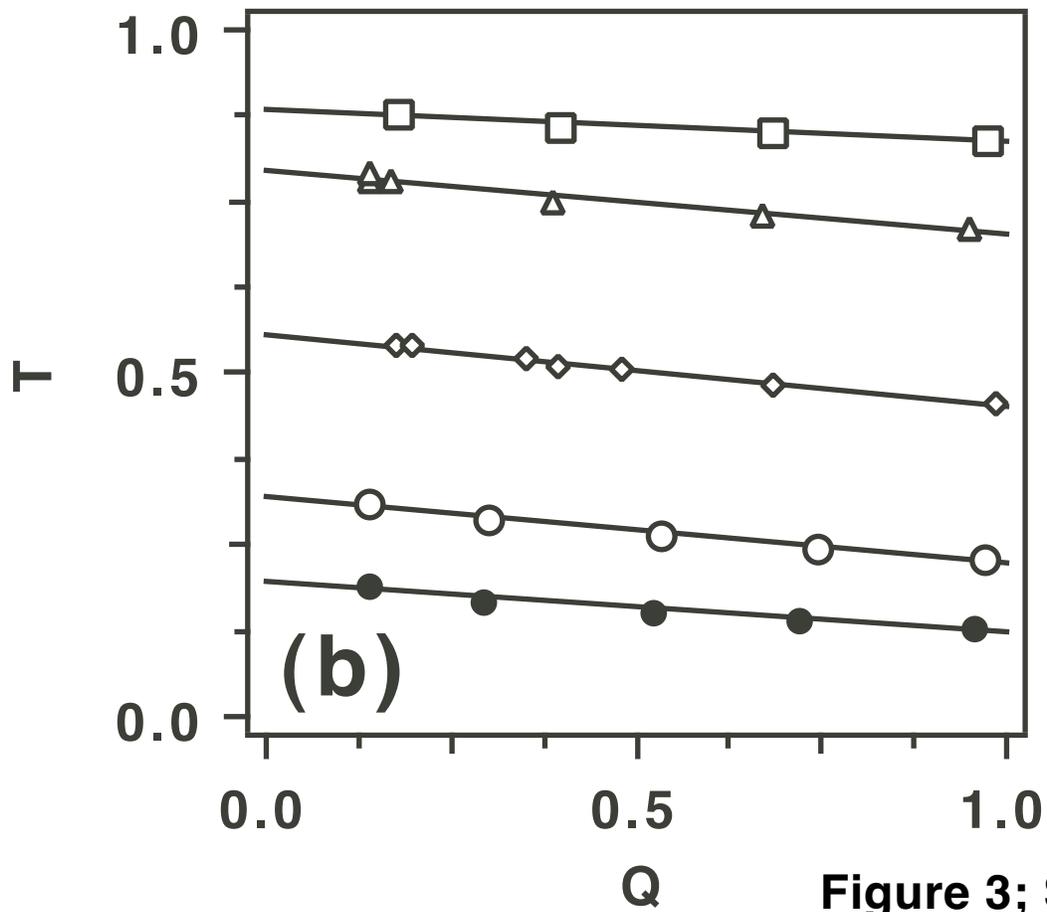

Figure 3; Sauer et al.

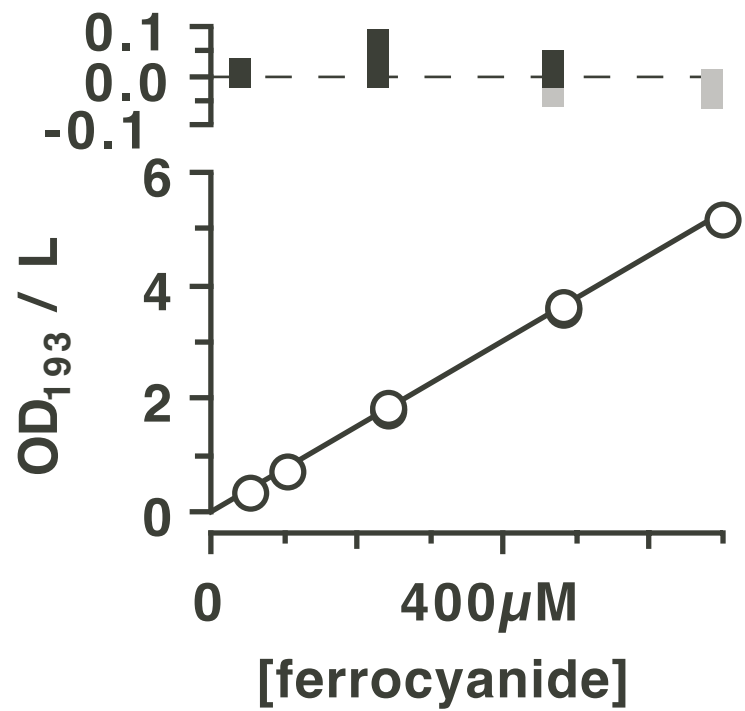

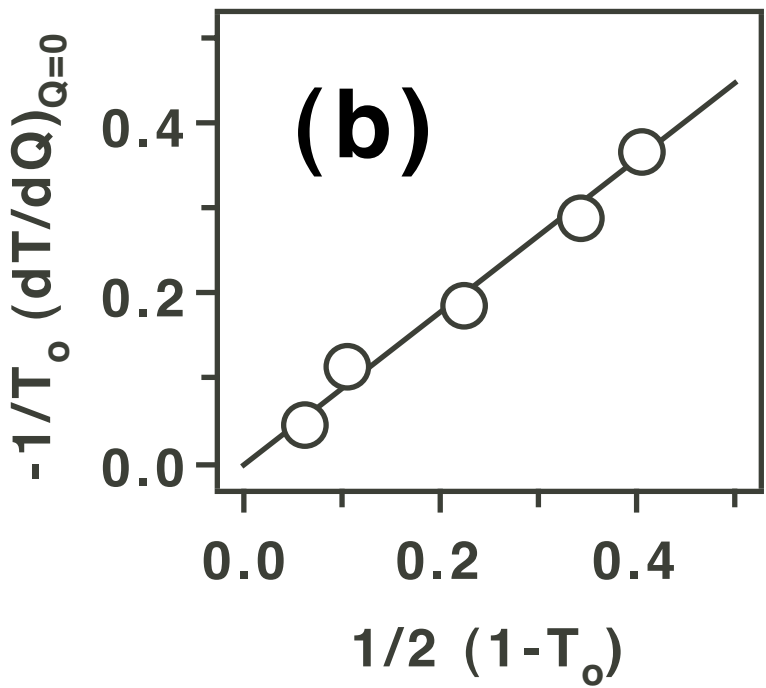

**Figure 4; Sauer et al.**

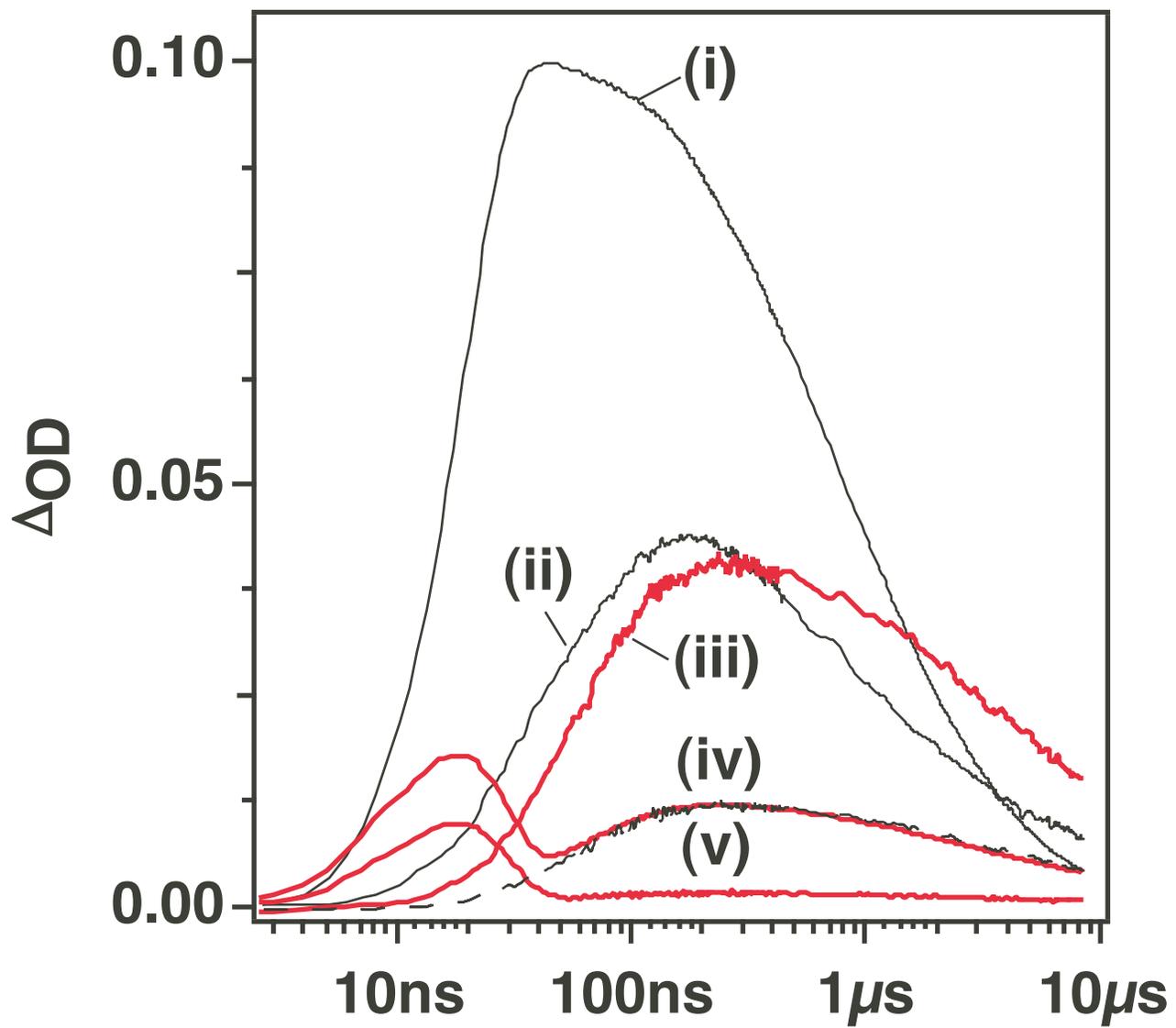

**Figure 5; Sauer et al.**

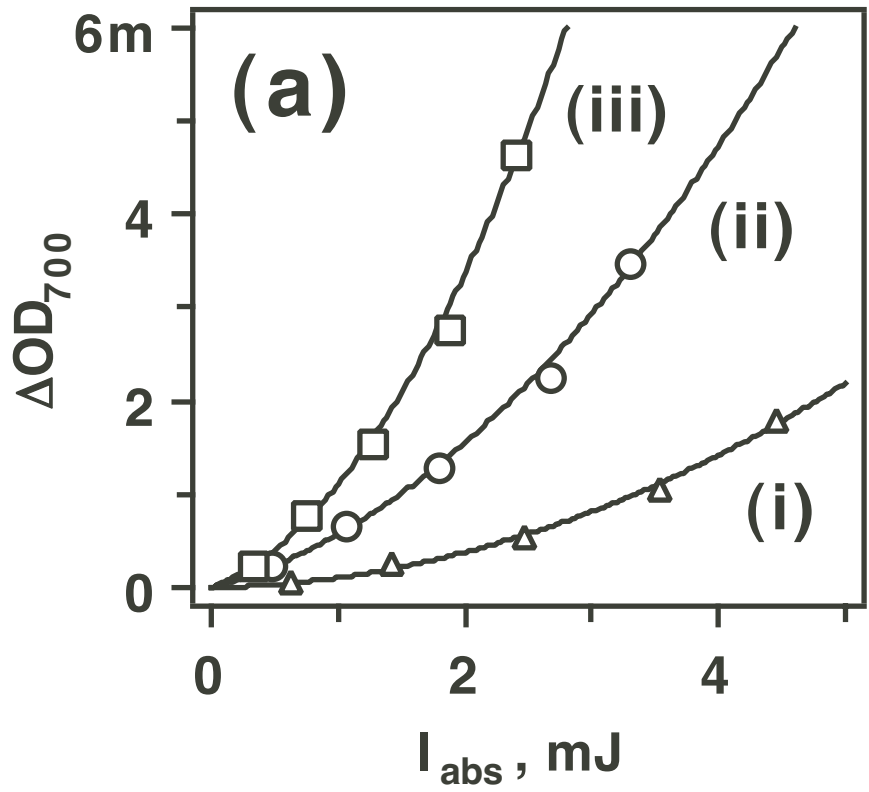
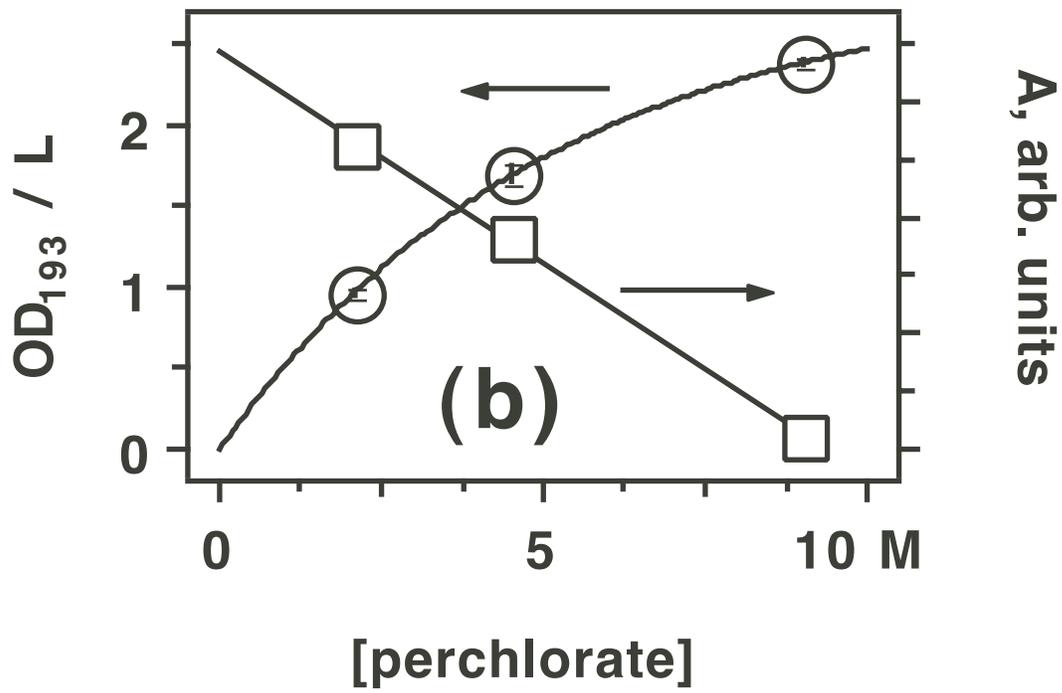

Figure 6; Sauer et al.

**SUPPLEMENTARY MATERIAL**                        **JP000000**

*Journal of Physical Chemistry A, Received \*, 2004*

**Supporting Information.**

(1S.) Captions to figures 1S to 12S.

**Fig. 1S.**

(a) Transient absorption kinetics (observed using 600 nm analyzing light) for the formation of carbonate radical in 248 nm photolysis of 24.3 mM *(thin red lines)* and 41.3 mM *(bold blue lines)* hydrogen peroxide in 1 M KHCO$_3$. Note the logarithmic time scale. Larger absorbance signals correspond to higher laser power. The photon fluences $N_{abs}$ are given in Fig. 2S(a). (b) *Black line:* Numerical simulation of photoconversion $\Phi$ of hydrogen peroxide to OH radicals as a function of dimensionless parameter $\phi N_{abs}/c_0 L$. For $\beta_{pr} = 0$, these two parameters would be equal (eq. (11), *blue line*), otherwise, there is a negative curvature (eq. (16)). Since OH radicals strongly absorb 248 nm light (an estimate of $\varepsilon_{248}(OH) \approx 508$ M$^{-1}$ cm$^{-1}$ was obtained from the laser light transmission data shown in Fig. 2S(b)), the photoconversion plot is curved. *Bold yellow line:* the least squares fit of the theoretical plot by the exponential dependence given by eq. (2).

**Fig. 2S.**

(a) *Yellow line and empty symbols, to the left:* The data of Fig. 1(a) replotted as the end-of-pulse concentration of hydroxyl radicals [OH] vs. $N_{abs}$. *Filled symbols, to the right:* the ratio of relative absorbances of 248 nm light by photogenerated OH radicals and H$_2$O$_2$ precursor for (i) 24.3 mM *(filled squares)* and (ii) 41.3 mM *(filled circles)* hydrogen peroxide. Due to 20:1 ratio of these molar absorptivities at the laser wavelength, relatively large fraction of 248 nm photons is absorbed by the hydroxyl radicals despite small (< 1%) photoconversion. (b) The transmission $T$ of 248 nm light as a function of dimensionless parameter $Q = \phi \beta I_0$ for (i) 24.3 mM and (ii) 41.3 mM hydrogen peroxide. The negative slope of these power dependencies indicates that $\beta_{pr} > \beta$ (sections 3 and 4.1.2).

**Fig. 3S.**

(a) Photoelectron kinetics (transient absorbance of 700 nm light) in 248 nm photoexcitation of aqueous sodium iodide (ca. 8.5 mJ of incident light). The iodide concentrations are indicated in the plot. At higher iodide concentration, the decay of the electron is faster, due to rapid cross recombination. (b) *Symbols:* End-of-pulse photoelectron absorbance in 248 nm photoexcitation of aqueous iodide vs. absorbed laser power. The solid line is a fit by eq. (2). *Empty squares:* fixed $I_0$=8.5 mJ, increasing iodide concentration (same as in (a), red to purple). *Empty circles:* fixed [I$^-$]=0.2 mM, increasing laser power.

**Fig. 4S.**

248 nm laser photolysis of (i) aqueous sodium hydrosulfide and (ii) aqueous sodium thiosulfate. Molar concentrations of the anions are 230 and 690 mM for HS$^-$ and 5.2 mM *(open circles)* and 9.5 mM *(open squares)* for $S_2O_3^{2-}$. (a) Photoelectron absorbance (700 nm) vs. absorbed laser power. (b) Optical density of these aqueous solutions at the laser wavelength vs. the anion concentration. (c) The transmission of laser light $T$ vs. incident laser power, $I_0$, for thiosulfate.

**Fig. 5S.**

Absorbance of 193 nm laser light by aqueous solutions (a,c) and power dependencies of the photoelectron absorbance (b,d) for several concentrations of sodium bicarbonate (a,b) and potassium hydroxide (c,d). The power dependence shown in (b) was obtained for $I_0$=6.9 mJ and four concentrations of bicarbonate (4.8, 9.1, 13, and 19 mM). In (d) the concentration of hydroxide was 0.4 mM *(open circles),* 0.79 mM *(open squares),* and 1.28 mM *(open diamonds).*

**Fig. 6S.**

Photoelectron absorbance (700 nm) vs. absorbed 193 nm laser power, for several anion photosystems. Increasing concentrations of these anions are indicated by color ascending in the spectral order, from red for the lowest concentration to purple for the highest concentration. Straight lines are least squares linear regressions of the initial sections of these power dependencies. (a) Power dependencies for (i) chloride (1.52 mM, *triangles*; 4.3 mM, *diamonds*; 7.9 mM, *squares*) and (ii) thiocyanate (38, 92, 200, and 305 µM). (b) Power dependencies for (i) sulfate (1.5, 4.3, 8.7, and 14.5 mM) and (ii) sulfite (190, 375, and 560 µM). (c) Power dependencies for (i) thiosulfate (8.7, 23, and 48 µM) and (ii) hydrosulfide (230, 450, and 680 µM).

**Fig. 7S.**

193 nm laser photolysis of (a,b) aqueous sodium bromide and (c,d) aqueous potassium iodide: 100 µM *(open squares),* 200 µM *(open circles),* and 300 µM *(open diamonds).* (a,c) Photoelectron absorbance vs. absorbed laser power. For iodide, a second series of measurements for 200 µM solution is shown by filled circles. (b,d) Laser light transmission $T$ vs. parameter $Q$.

**Fig. 8S.**

(a) Ratio $\left[CO_3^{2-}\right]/\left[OH^-\right]$ and (b) molar concentrations of hydroxide, bicarbonate, *(to the left)* and carbonate anions *(to the right)* vs. molar concentration of $Na_2CO_3$. No corrections for non-unity activity coefficient were made; ionization constants for carbonic acid (for infinite dilution) were taken from ref. [48].

**Fig. 9S.**

Same as Fig. 8S(b), for bicarbonate solutions (see the legend in the plot).

**Fig. 10S.**

The data of Fig. 3(a) replotted as photoconversion $\Phi$ of the ferrocyanide anion vs. (a) $N_{abs}$ and (b) dimensionless parameter $\phi N_{abs}/c_0 L$. Compare the latter plot with the simulated photoconversion dependencies given in Fig. 11S(a).

**Fig. 11S.**

Simulation of the data of Figs. 3 and 10S using formulae given in section 3. See Tables 1 and 3 for the simulation parameters. Note that photoconversion plots for different ferrocyanide concentrations shown in (a) exhibit much less divergence for different ferrocyanide concentrations than the experimental plots shown in Fig. 10S(b). As explained in section 4.2.2, the likely cause of this divergence is ion pairing that occurs in concentrated solutions of ferrocyanide.

**Fig. 12S.**

The same data as in Figs. 3 and 4, for 248 nm photolysis of the ferrocyanide. The concentrations of $K_4Fe(CN)_6$ are 0.4 mM *(open circles),* 0.8 mM *(open squares),* and 1.2 mM *(open diamonds).* (a) Transmission $T$ of the 248 nm light vs. $Q$. (b) Optical density $OD_{248}$ of the laser light for $Q \to 0$ (extrapolated from the plots in (a)) vs. ferrocyanide concentration. (c) Electron absorbance vs. laser power. The solid line is a least squares fit obtained using eq. (2); the straight line is the initial slope of this power dependence.

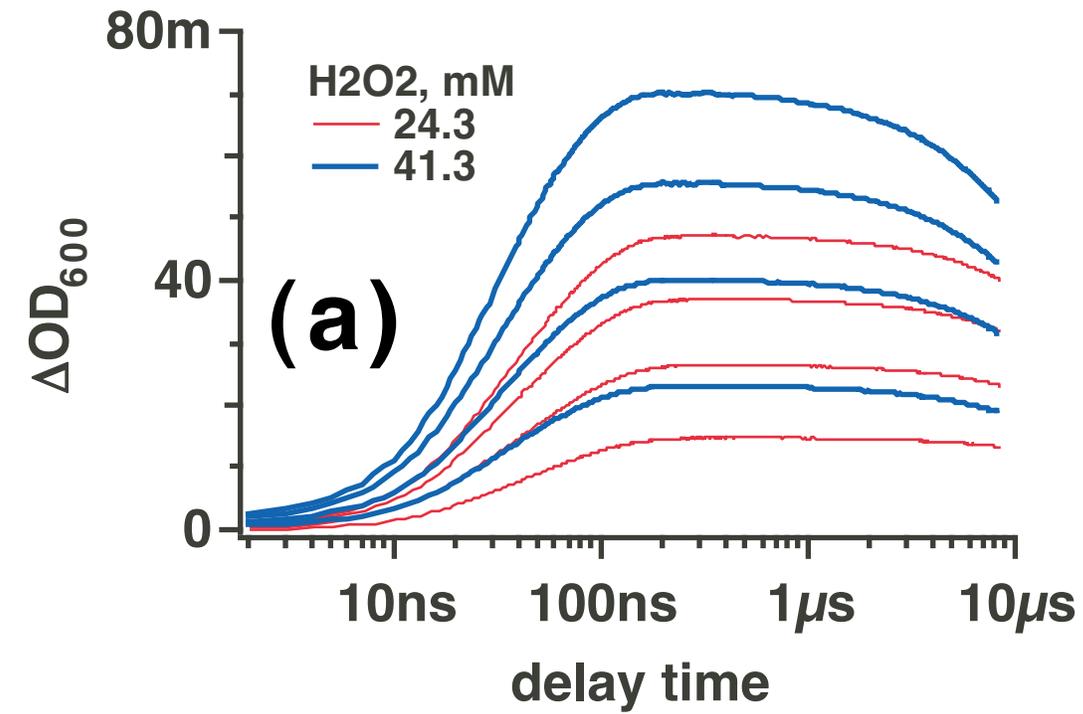

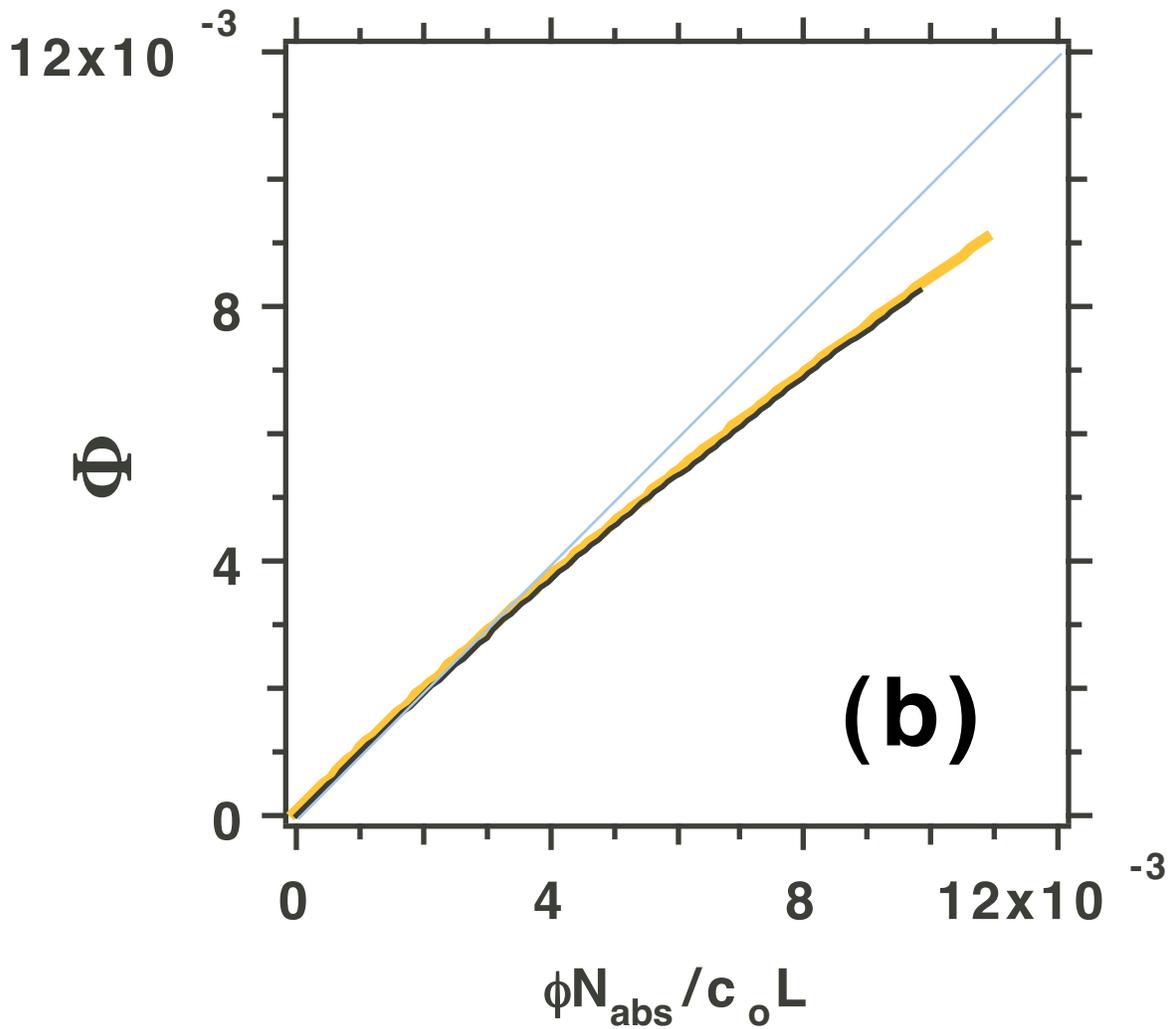

**Figure 1S; Sauer et al.**

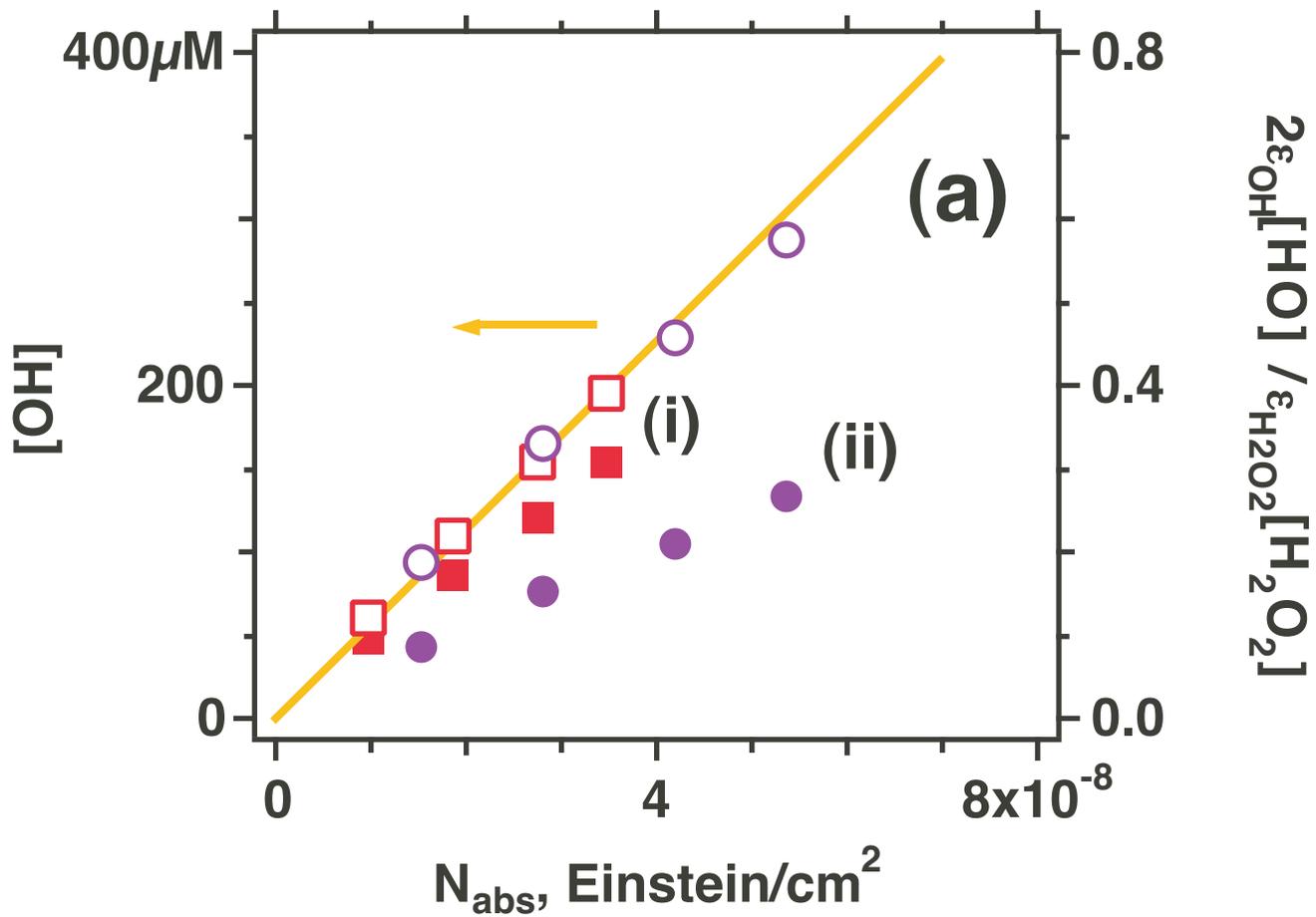
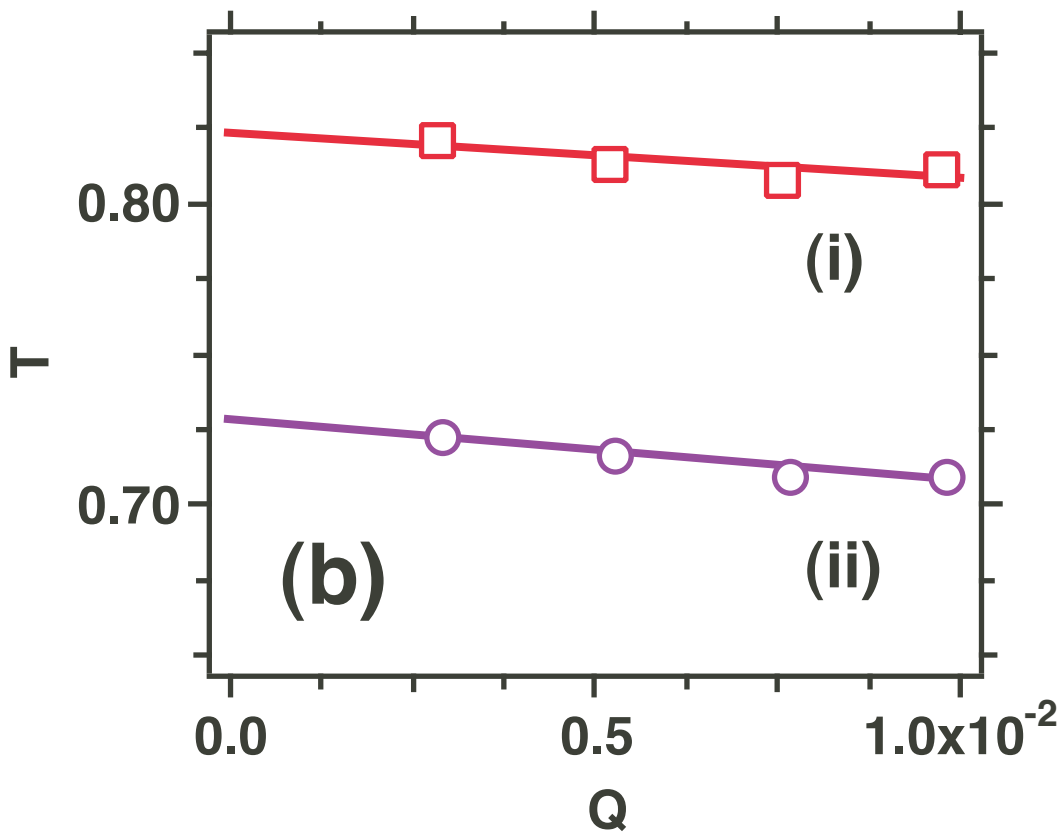

Figure 2S; Sauer et al.

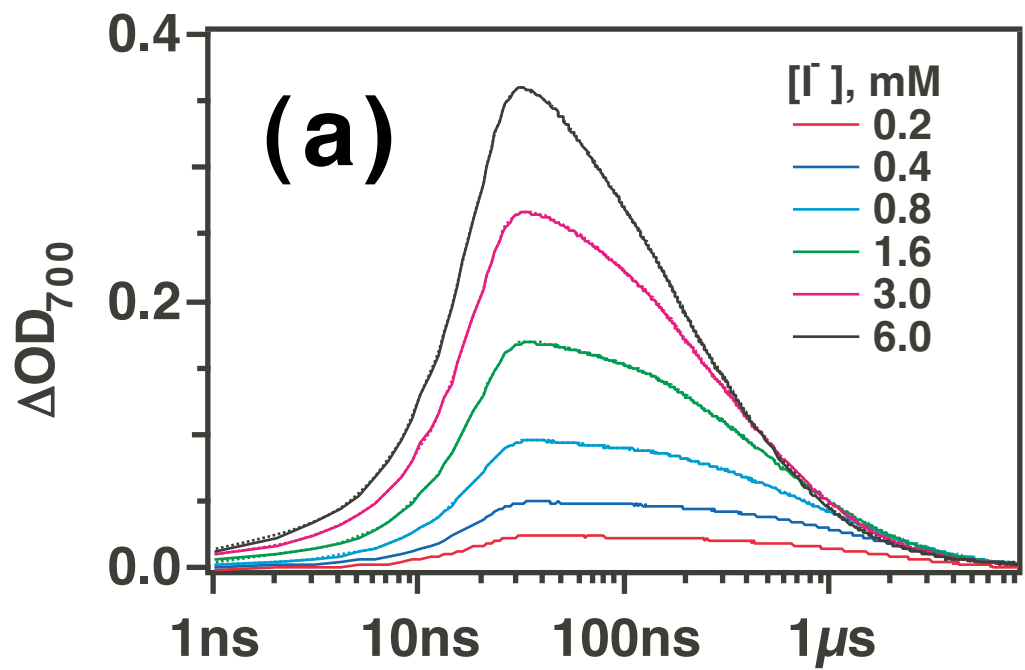

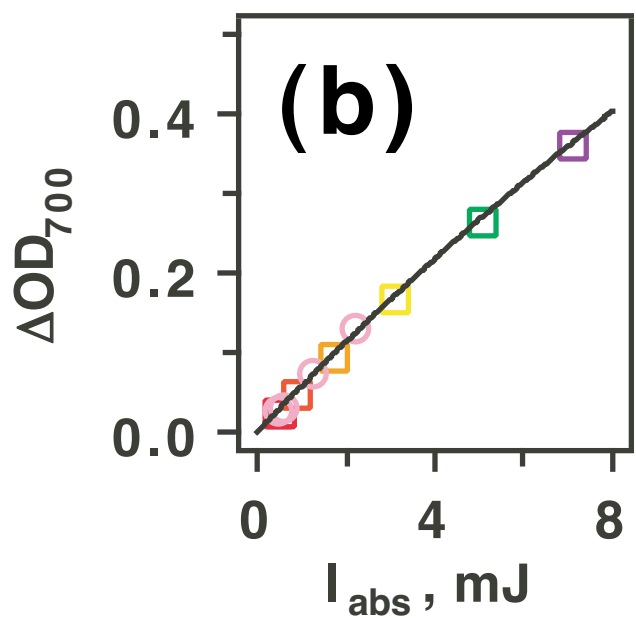

**Figure 3S; Sauer et al.**

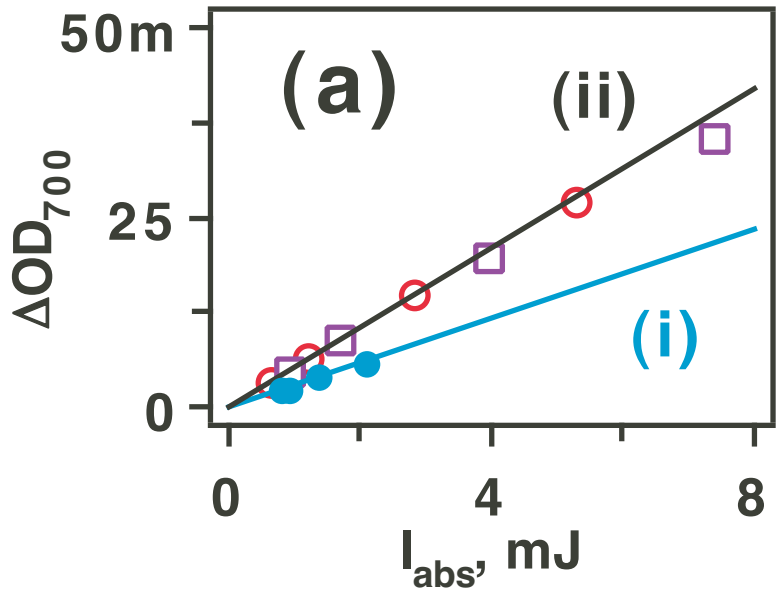

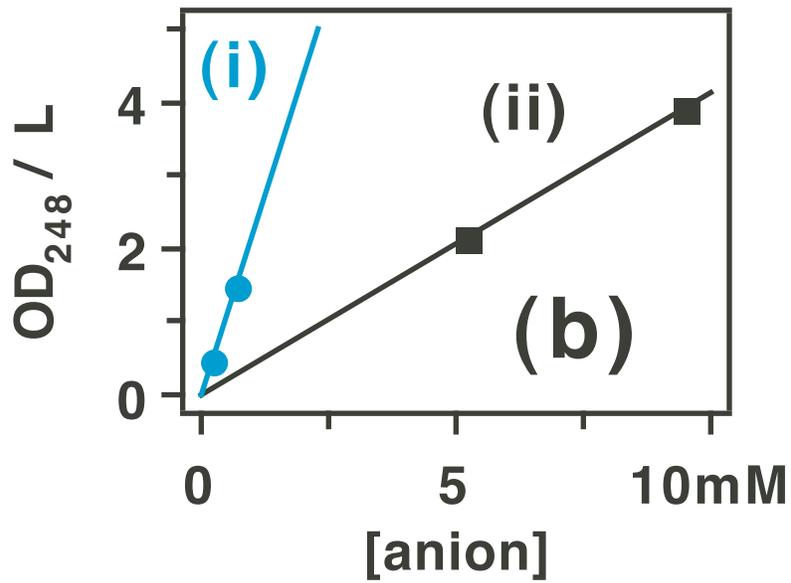

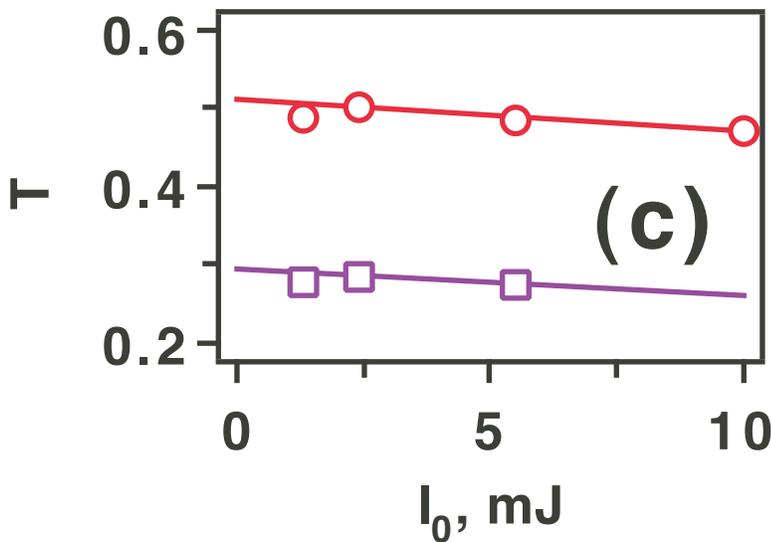

**Figure 4S; Sauer et al.**

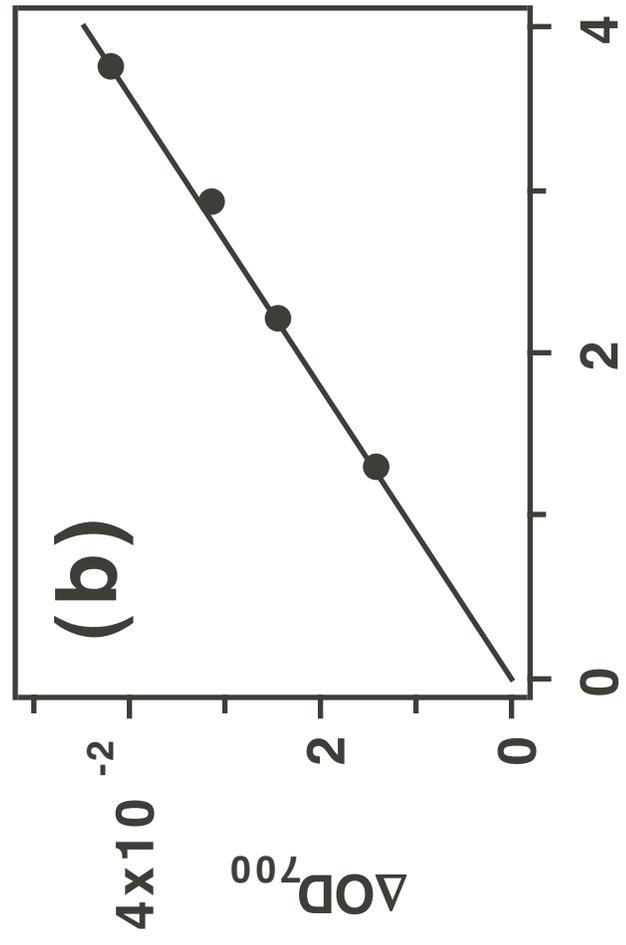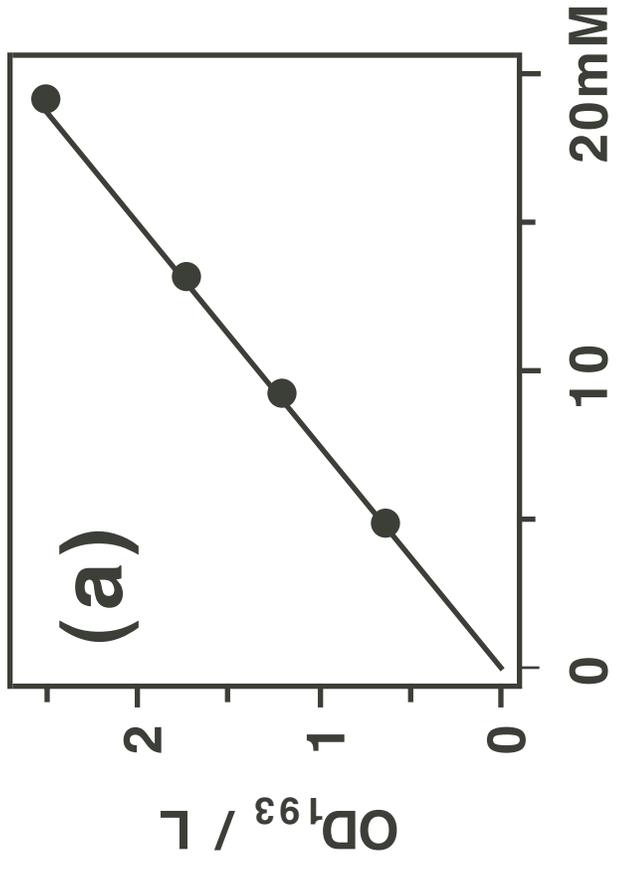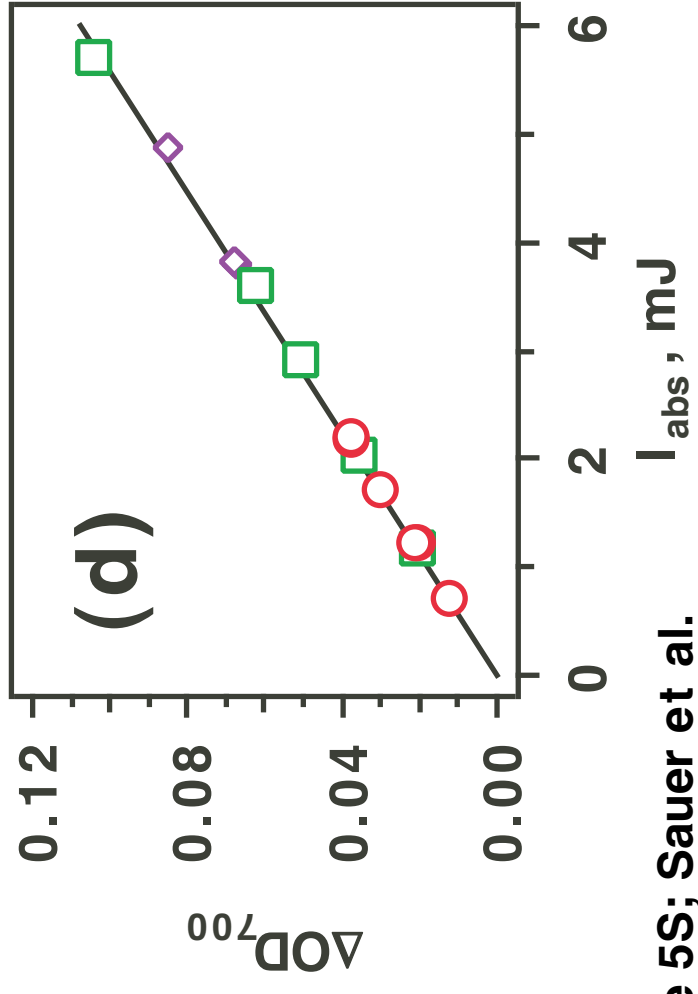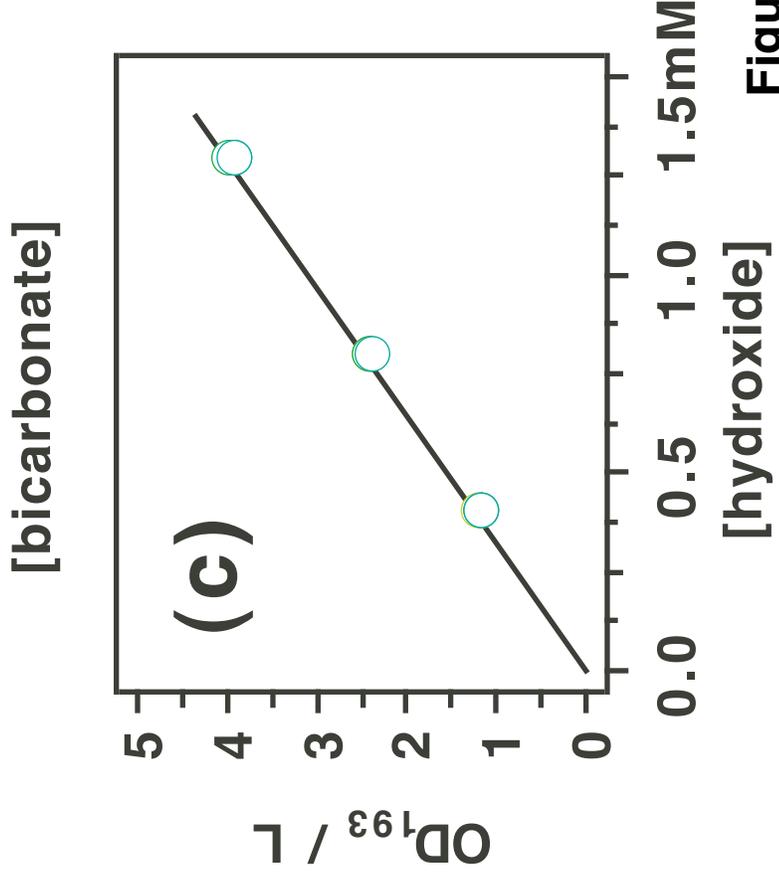

Figure 5S; Sauer et al.

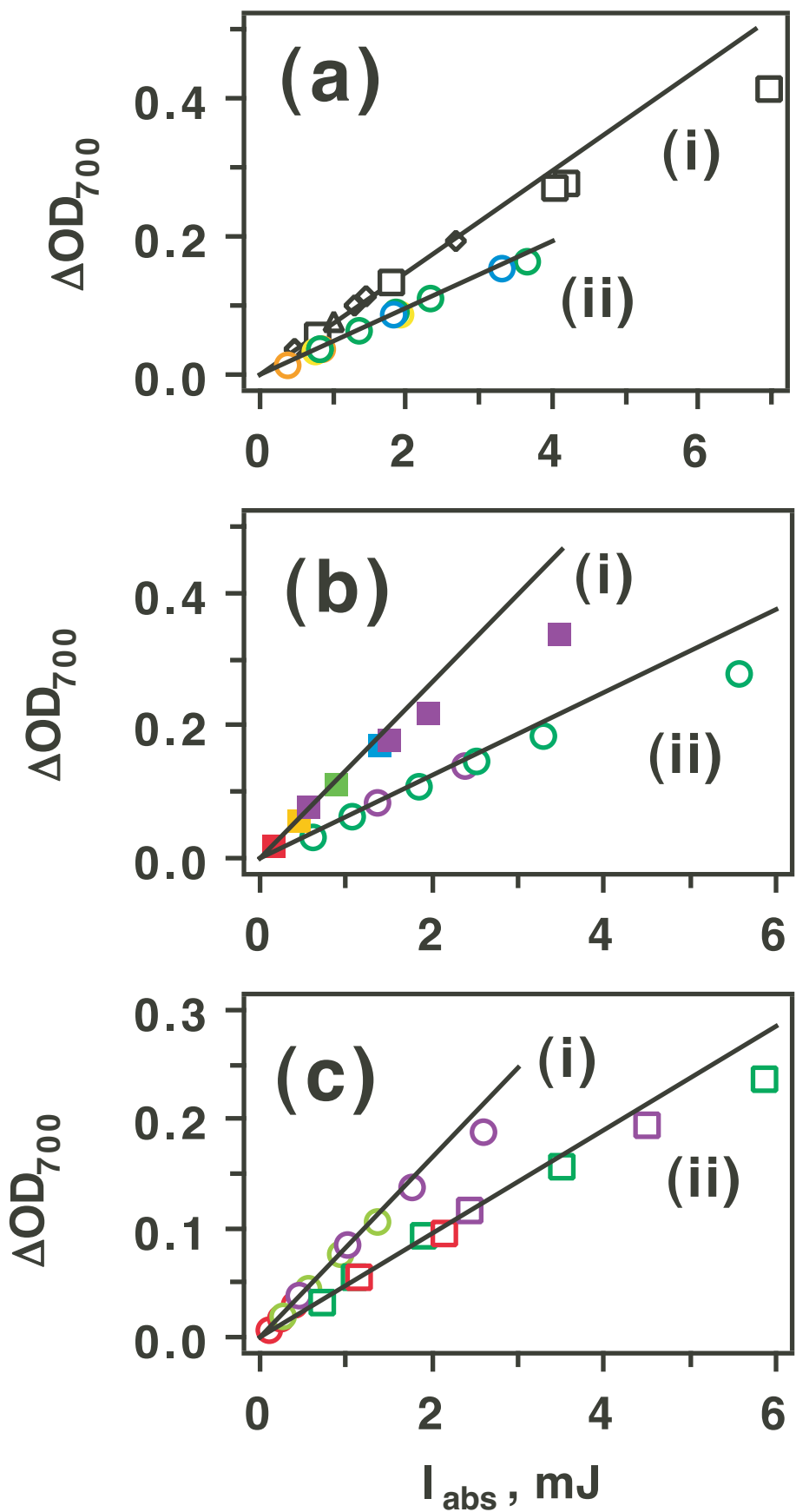

Figure 6S; Sauer et al.

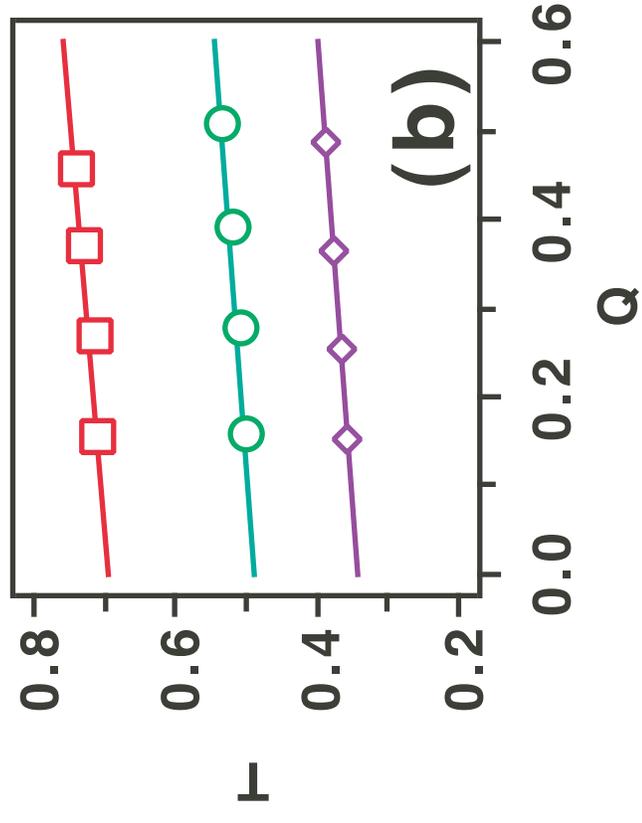
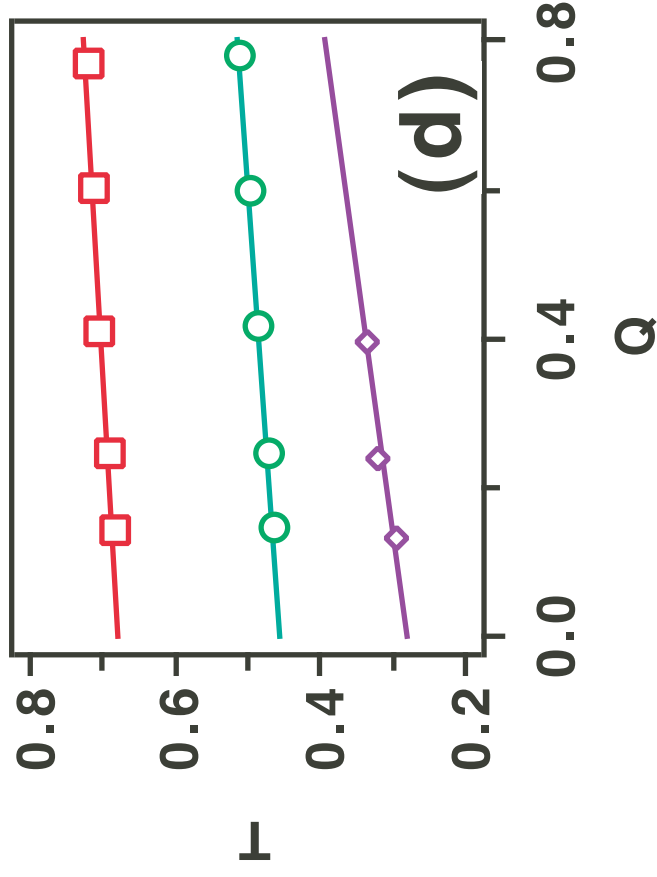
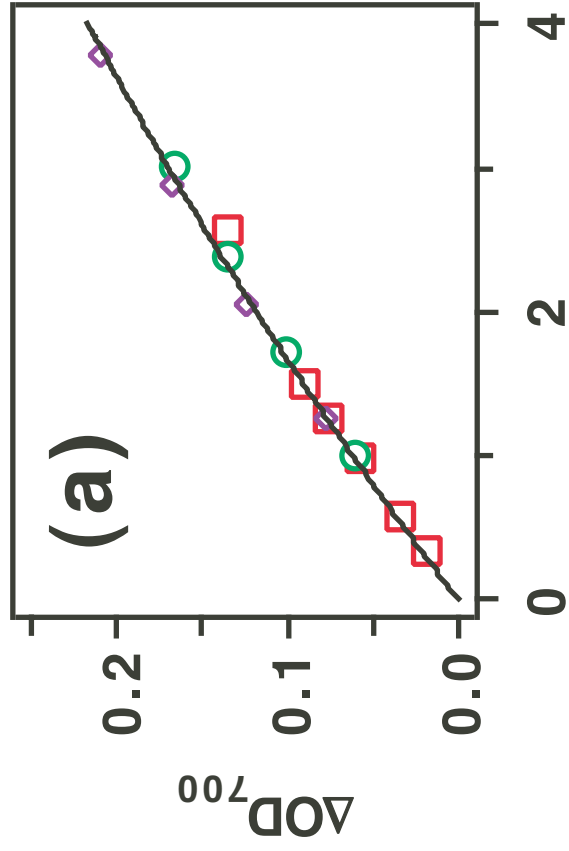
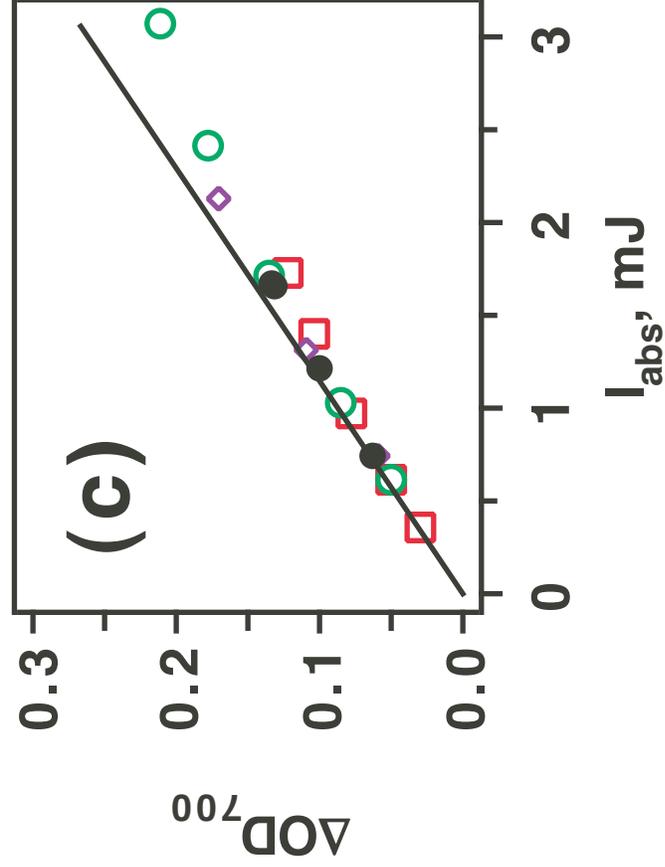

**Figure 7S; Sauer et al.**

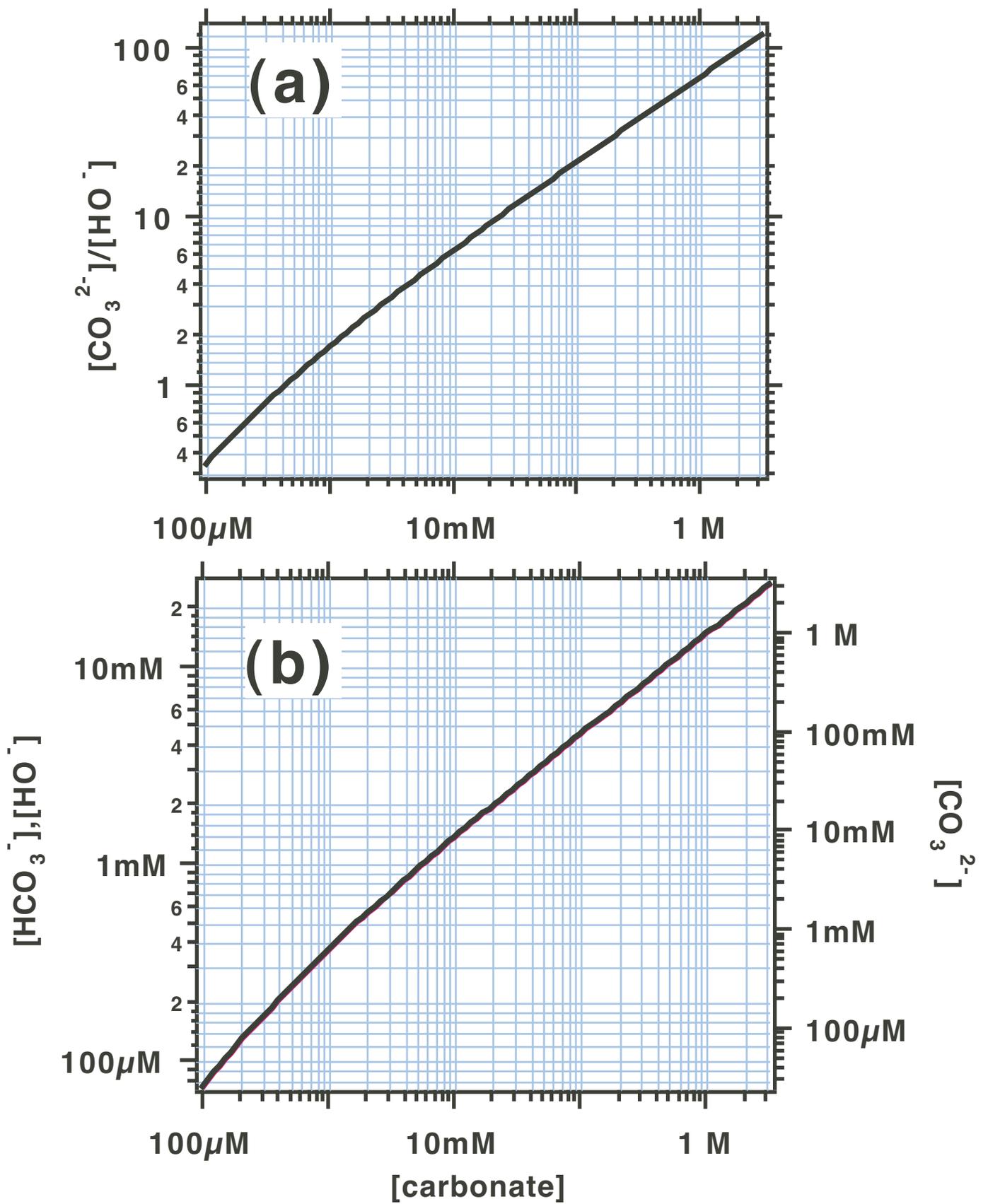

**Figure 8S; Sauer et al.**

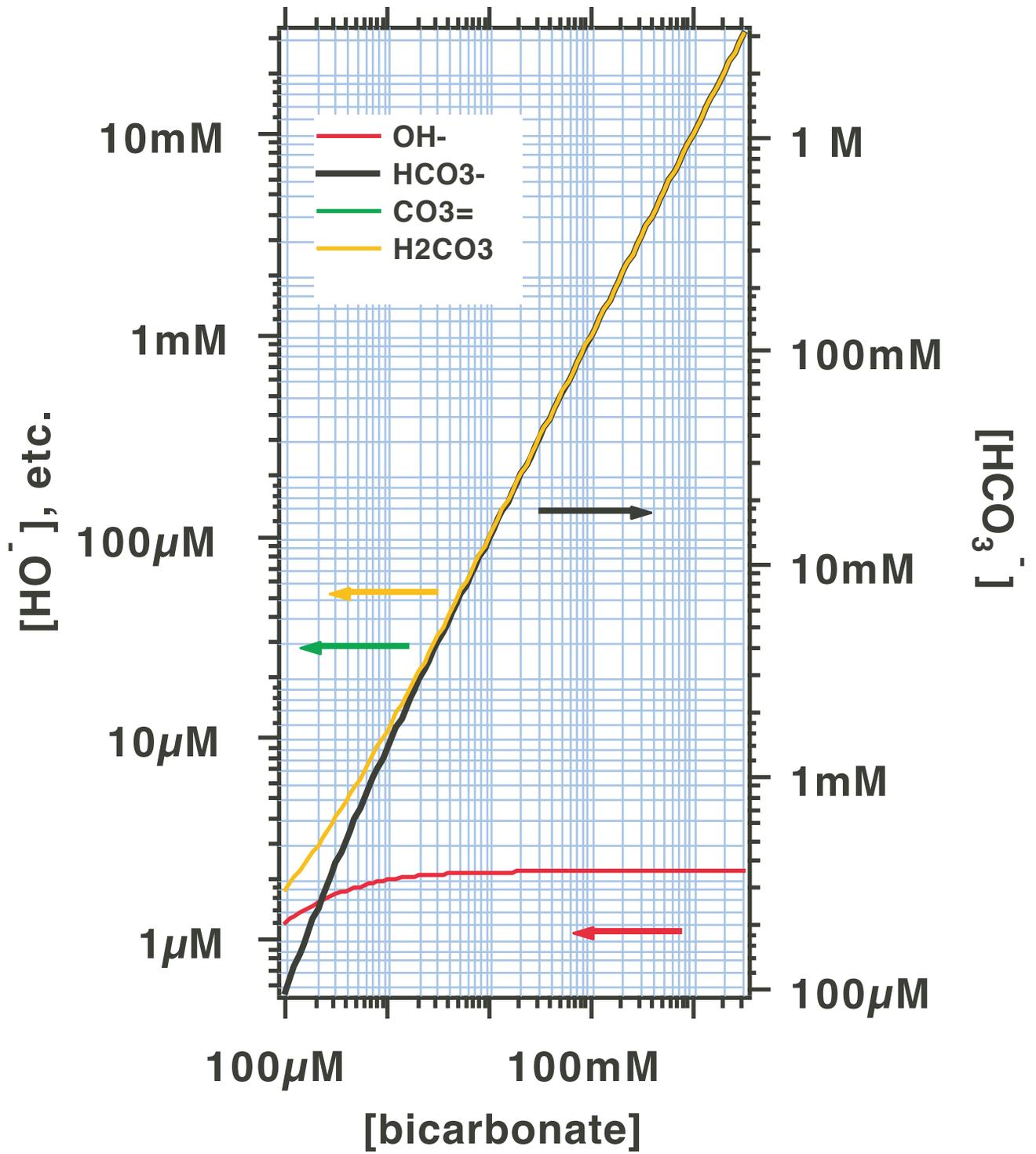

Figure 9S; Sauer et al.

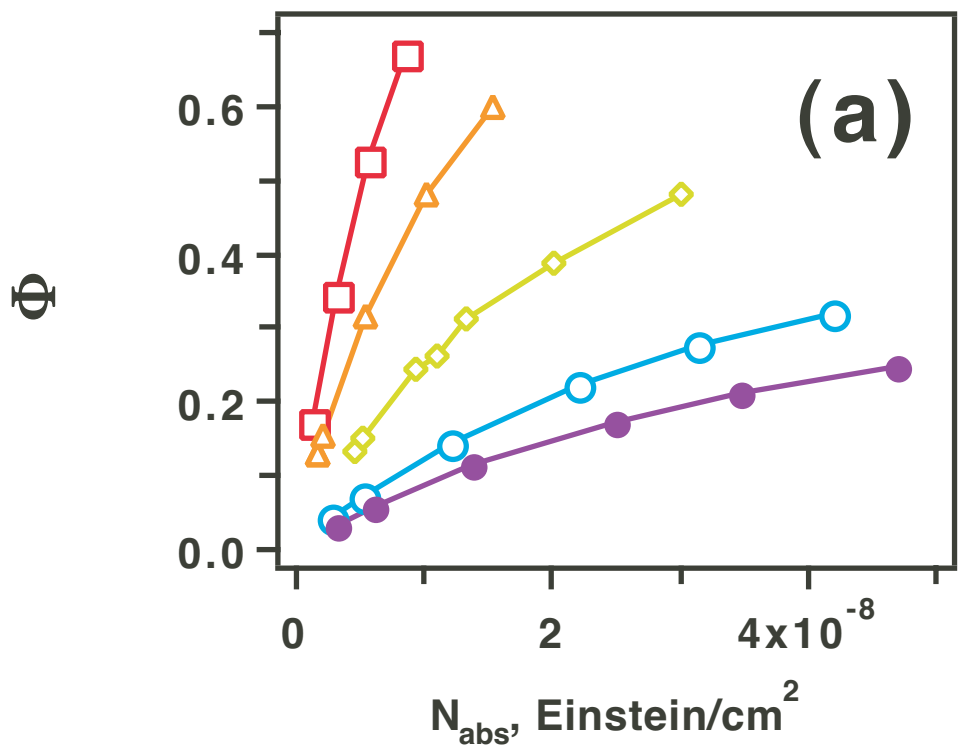

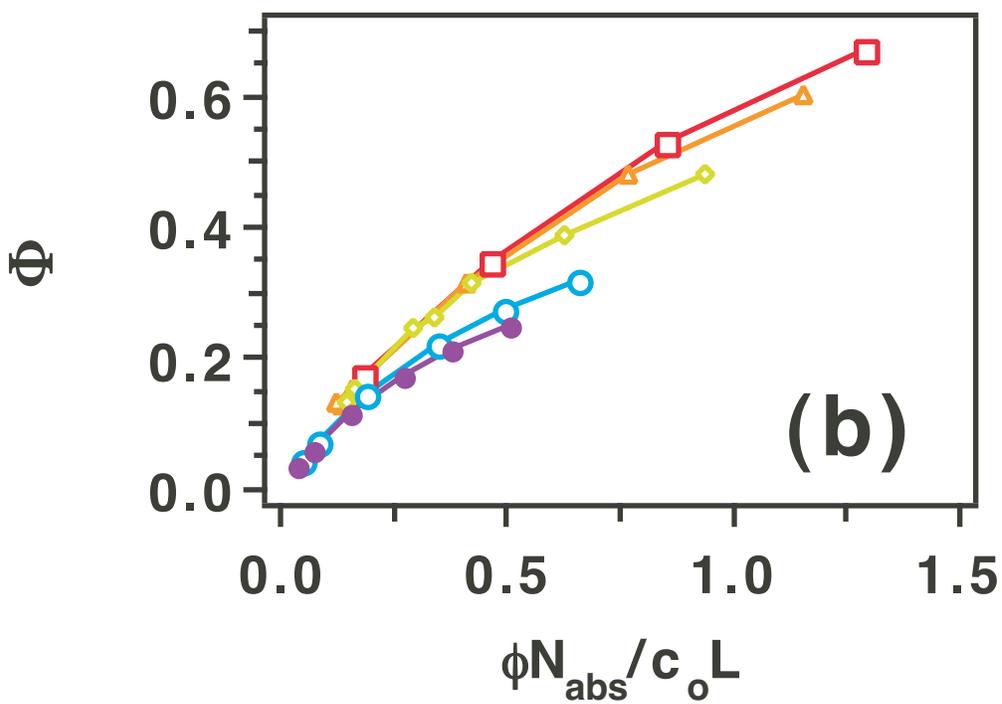

**Figure 10S; Sauer et al.**

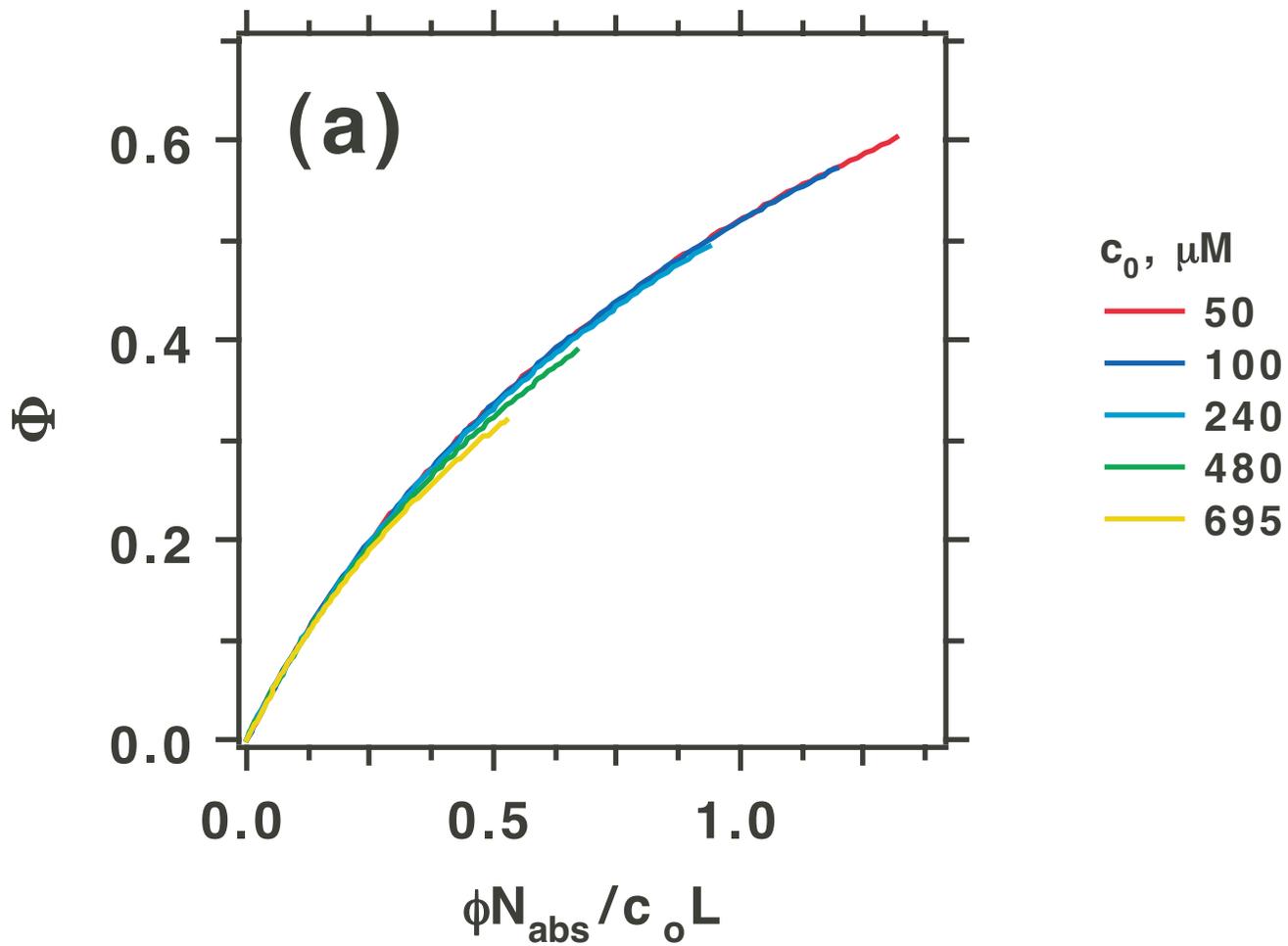

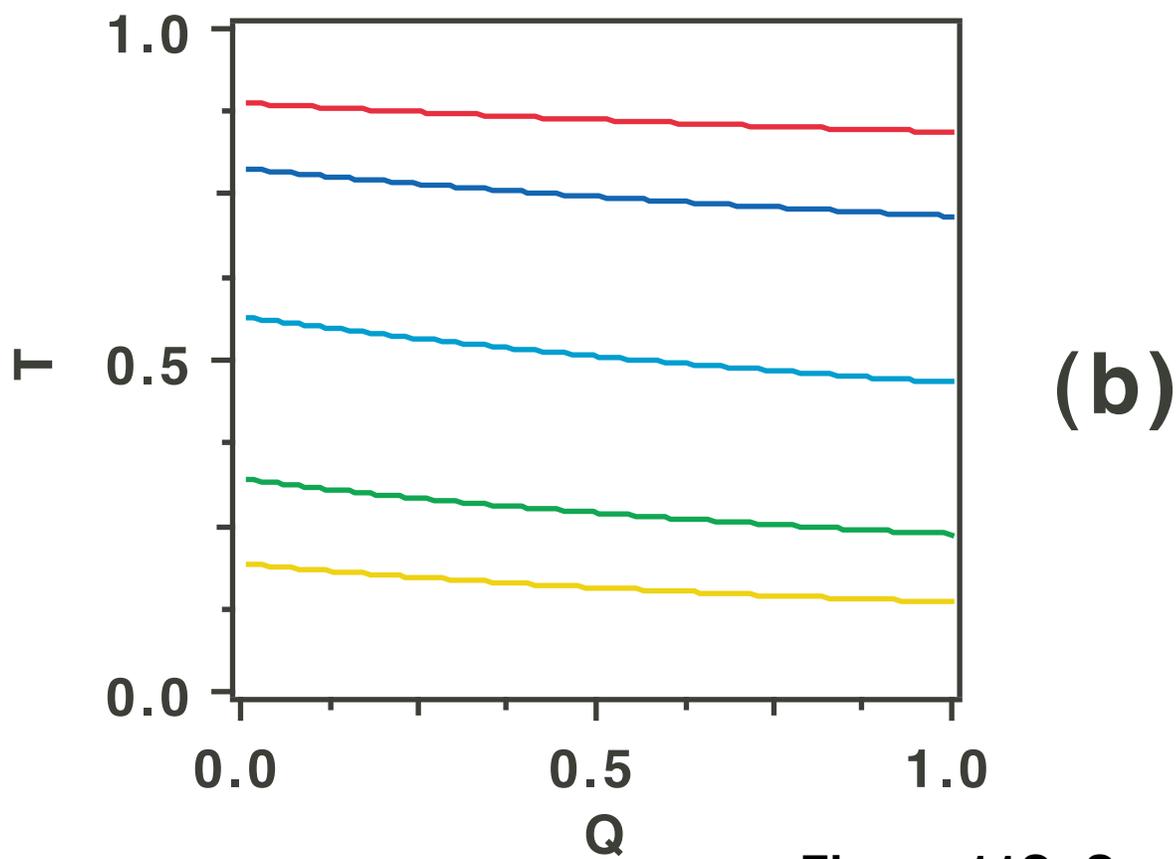

**Figure 11S; Sauer et al.**

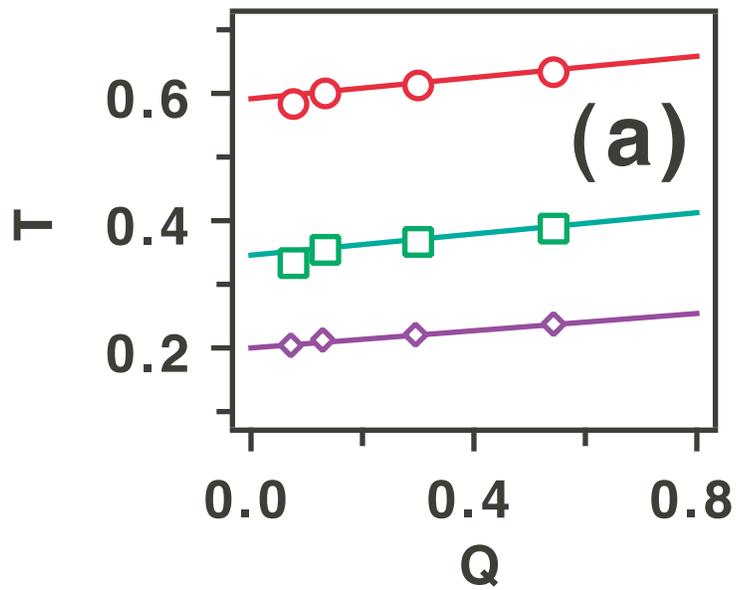

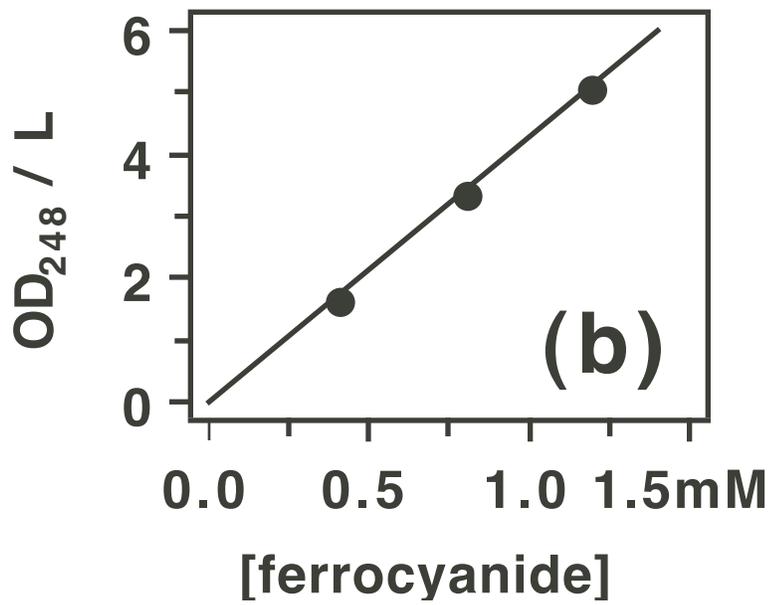

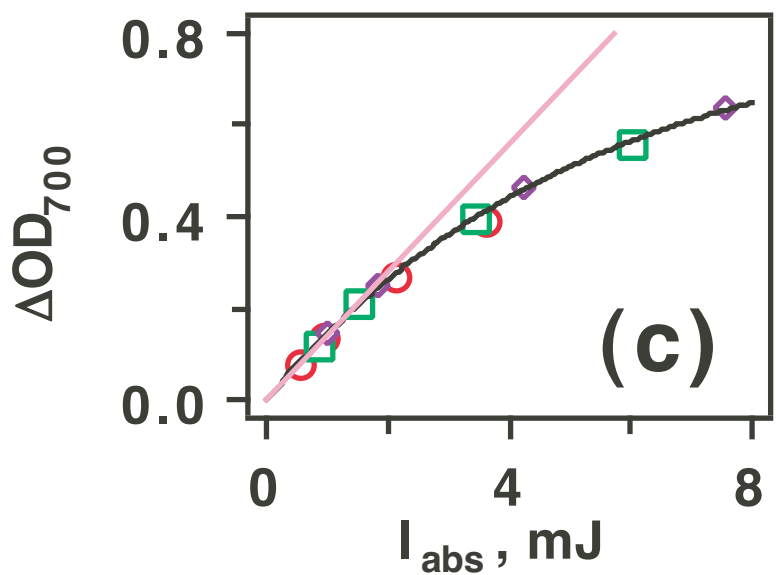

**Figure 12S; Sauer et al.**